\documentclass[twocolumn]{aastex61}

\expandafter\def\expandafter\normalsize\expandafter{%
    \normalsize
    \setlength\abovedisplayskip{10pt}
    \setlength\belowdisplayskip{10pt}
    \setlength\abovedisplayshortskip{10pt}
    \setlength\belowdisplayshortskip{10pt}
}
\received{XXXX}
\revised{XXXX}
\accepted{XXXX}

\usepackage{natbib}
\usepackage{comment}
\usepackage{amsmath}
\usepackage{float}
\usepackage{csquotes}

\begin{document}

\title{\textit{K2} Ultracool Dwarfs Survey. III. White Light Flares are Ubiquitous in M6-L0 Dwarfs}

\author{Rishi R. Paudel, John E. Gizis, D. J. Mullan}
\affiliation{Department of Physics and Astronomy, University of Delaware, Newark, DE, 19716, USA}

\author{Sarah J. Schmidt}
\affiliation{Leibniz-Institute for Astrophysics Potsdam (AIP), An der Sternwarte 16, 14482, Potsdam, Germany}
\author{Adam J. Burgasser}
\affiliation{Center for Astrophysics and Space Science, University of California San Diego, La Jolla, CA 92093, USA}
\author{Peter K. G. Williams}
\affiliation{Harvard-Smithsonian Center for Astrophysics, 60 Garden Street, Cambridge, MA 02138, USA}
\author{Edo Berger}
\affiliation{Harvard-Smithsonian Center for Astrophysics, 60 Garden Street, Cambridge, MA 02138, USA}

\begin{abstract}
We report the white light flare rates for 10 ultracool dwarfs (UCDs) using \textit{Kepler K2} short cadence data. Among our sample stars, two have spectral type M6, three are M7, three are M8 and two are L0. Most of our targets are old low mass stars. We identify a total of 283 flares in all of the stars in our sample, with \textit{Kepler} energies in the range log \textit{E$_{Kp}$} $\sim$(29 - 33.5) erg. Using the maximum-likelihood method of line fitting, we find that the flare frequency distribution (FFD) for each star in our sample follows a power law with slope -$\alpha$ in range -(1.3-2.0). We find that cooler objects tend to have shallower slopes. For some of our targets, the FFD follows either a broken power law, or a power law with an exponential cutoff. For the L0 dwarf 2MASS J12321827-0951502, we find a very shallow slope (-$\alpha$ $=$ -1.3) in the \textit{Kepler} energy range (0.82-130)$\times$10$^{30}$ erg: this L0 dwarf has flare rates which are comparable to the rates of high energy flares in stars of earlier spectral types. In addition, we report photometry of two superflares: one on the L0 dwarf 2MASS J12321827-0951502 and another on the M7 dwarf 2MASS J08352366+1029318. In case of 2MASS J12321827-0951502, we report a flare brightening by a factor of $\sim$144 relative to the quiescent photospheric level. Likewise, for 2MASS J08352366+1029318, we report a flare brightening by a factor of $\sim$60 relative to the quiescent photospheric level. These two superflares have bolometric (UV/optical/infrared) energies 3.6 $\times$ 10$^{33}$ erg and 8.9 $\times$ 10$^{33}$ erg respectively, while the FWHM time scales are very short, $\sim$2 minutes. We find that the M8 star TRAPPIST-1 is more active than the M8.5 dwarf: 2M03264453+1919309, but less active than another M8 dwarf (2M12215066-0843197). 

\end{abstract}



\section{INTRODUCTION} \label{sec:introduction}

Ultracool dwarfs (hereafter UCDs) are stellar or substellar objects with effective temperatures no more than 2700K \citep{1999AJ....118.2466M,1999ApJ...519..802K}. Understanding the nature of magnetic dynamos in UCDs has been very challenging. The UCDs show deviations from age, rotation and activity relations which are seen in early and mid-M dwarfs. The rotation rate plays a significant role in shaping the magnetic dynamo of early and mid-M dwarfs. When the stars are young, their rapid rotation rates empower strong magnetic dynamos. As they evolve, magnetic braking slows the rotation, which in turn decreases the magnetic activity \citep{2009ARA&A..47..333D, 2005stam.book.....G, 2005ApJ...622..653T}. The UCDs (which include both young brown dwarfs and old low-mass stars), despite being rapid rotators, display a poor relation between rotation and magnetic activity. Some of the usual indicators of activity (X-ray and H$_{\alpha}$ emission) weaken significantly among the UCDs \citep{2000AJ....120.1085G,2010ApJ...709..332B,2015AJ....149..158S}. This could be due to a number of factors, including cool atmospheres which have reduced amounts of ionized gas \citep{2002ApJ...571..469M}, or atmospheres which are undergoing centrifugal coronal stripping \citep{2000MNRAS.318.1217J,1999A&A...346..883J,2008ApJ...676.1307B}. On the other hand, observations of radio emission show that strong magnetic fields exist in UCDs \citep{2012ApJ...747L..22R,2015ApJ...808..189W,2016ApJ...830...85R}. One interpretation of these data is that turbulent dynamos in UCDs may be producing both large and small scale magnetic fields \citep{2010A&A...522A..13R, 2015ApJ...813L..31Y}. Alternatively, X-rays and H$_{\alpha}$ emission may be powered by fast magnetic reconnection, whereas radio emission may be generated by electrons which emerge from slow magnetic reconnection \citep{2010ApJ...721.1034M}.  
\\
\\
Stellar flares are transient events which are caused by sudden releases of magnetic energy in the upper atmosphere of the star. During this process, energy which was previously stored in magnetic form is converted (by reconnection) in part to kinetic energy of electrons and ions, in part to bulk flow of ejected matter (coronal mass ejections: CMEs) \citep{2010ApJ...721.1034M,2010ARA&A..48..241B}, and in part to thermal energy. White light flares (hereafter WLFs) are assumed to be produced when non thermal electrons accelerated after reconnection hit a cold thick target in lower chromosphere or upper photosphere. The precipitated electrons cause the formation of hot \enquote{chromospheric condensations} which emit white light continuum: the continuum has a wavelength dependence in visible photons which approximates that of a blackbody with a temperature of order 10$^{4}$ K (\cite{2015SoPh..290.3487K} and references therein). During a WLF, a faint star can become significantly brighter in optical light,
by as much as several magnitudes. Two huge WLFs on UCDs will be discussed later in this paper. Estimates of surface areas of flares show that the  WLFs in UCDs cover larger fractional areas of the surface than do flares on bright stars such as the Sun \citep{2010ApJ...714L..98K,2011AJ....141...50W}. 
\cite{2017ApJ...846...75P} suggest that flares on brown dwarfs could result from planet like auroral emissions produced by large-scale magnetospheric currents. 
\\
\\
The discovery of the TRAPPIST-1 planetary system \citep{2016Natur.533..221G,2017Natur.542..456G,2017NatAs...1E.129L} around an M8.0 dwarf suggests that more such planetary systems around other M stars or possibly brown dwarfs may be discovered. Thus, there will be increased chances of finding Earth like planets in the habitable zones (HZ) of such stars. To know the habitable conditions of such planets, it will be important to know flare rates on the host stars: UV photons from huge flares may have significant impact on the chemical evolution of the atmospheres of the planets \citep{2010AsBio..10..751S,2014P&SS...98...66G,2016MNRAS.455.3110A,2016MNRAS.459.4088O}. In this regard, studies of the WLF rates of UCDs will contribute to the conclusions. \\
\\
The \textit{Kepler} mission \citep{2010ApJ...713L..79K} was originally aimed at finding more Earth sized planets. But it has also turned out to be useful for studying stellar properties, including WLF rates, astroseismology, etc. WLF rates of several early-M and mid-M dwarfs were estimated using \textit{Kepler} data \citep{2013MNRAS.434.2451R,2013A&A...555A.108M,2014ApJ...797..121H,2016ApJ...829...23D}. The occurence of WLFs on L dwarfs and young brown dwarfs  \citep{2013ApJ...779..172G,2016ApJ...828L..22S,2017ApJ...838...22G,2017ApJ...845...33G}  is a proof that WLFs are common in some UCDs. In the present paper, we continue our monitoring of various UCDs using \textit{Kepler K2 } \citep{2014PASP..126..398H} data, with a goal of studying the WLF rates of various UCDs. Our targets, which include mainly late-M dwarfs and early L dwarfs, were monitored by \textit{Kepler K2} mission during Campaigns 3, 4, 5, 6 and 12. Some targets in our sample have spectral type M6. We consider the objects with spectral types $\geq$M6 as UCDs in this paper. The results of work presented here is an important step that in future works will be helpful to understand different flare properties (e.g. flare energy, duration, rate, etc) in UCDs and how these properties depend on spectral type, age, mass, etc. In cases where rotation periods and ages of targets are known, our study may shed light on the rotation-age-activity relationships in UCDs. We also include in this paper our own analysis of TRAPPIST-1 flares which were previously discussed by \cite{2017ApJ...841..124V} and \cite{2017RNAAS...1....2D}.  \\
\\
In this paper, we discuss flare properties in the context of the flare frequency distribution (FFD). For each star, the FFD is assumed to be fitted (over a range of energies) by a power law  \citep{1972Ap&SS..19...75G,1976ApJS...30...85L}:
\begin{equation} \label{eq:cu_power_law}
log\tilde{\nu} = \alpha_{o} - \beta logE
\end{equation}
%
where \textit{$\tilde{\nu}$} is cumulative (or integrated) flare frequency, i.e. the number of flares with energies of $\geq$ \textit{E} which were detected per unit observation time. The constant $\alpha_{o}$ represents intercept at zero energy, and the constant $\beta$ represents slope of the FFD. In stars which are found to have spectral index $\beta$ $>$ 1, the weakest flares contribute most to the total energy emitted by flares. In the stars with $\beta$ $<$ 1, the strongest flares contribute most to the total energy emitted by flares. Here, the phrase \enquote{total energy emitted by flares} refers to  energy of all flares which were detected during a given observation time. The FFD can also be expressed in terms of differential form as:
\begin{equation} \label{eq:diff_energy_power_law}
dN \propto E^{-\alpha}dE
\end{equation}
%
where \textit{dN} is the number of flares having energies in the range \textit{E} and \textit{E+dE}. The indices in Eq. \ref{eq:cu_power_law} and Eq. \ref{eq:diff_energy_power_law} are related as $\alpha$ = $\beta$ + 1. Many studies have been done to compute the FFD in the Sun and in early-M and mid-M dwarfs. \cite{1987SvA....31..231K} reported the value of spectral index $\beta$ to be $\sim$ 0.80 for the energy distribution of 15000 solar flares observed during 1978-79. \cite{2011PhDT.......144H} calculated $\beta$ = 0.73$\pm$0.1 in the \textit{U}-band energy range 10$^{27.94}$ $\leq$ \textit{E$_{U}$} $\leq$ 10$^{30.60}$ erg for four M6-M8 dwarfs. Likewise, using a simple linear fit \citep{2017ApJ...845...33G} reported $\beta$ = 0.59 ±$\pm$0.09 in energy range 10$^{31}$ erg to 2$\times$10$^{32}$ erg for a field L1 dwarf WISEP J190648.47+401106.8 (hereafter W1906+40)  and $\beta$ = 0.66$\pm$0.04 in energy range  4 $\times$ 10$^{31}$ erg to 1.1 $\times$ 10$^{33}$ for a 24 Myr brown dwarf  2MASS J03350208+2342356 (hereafter 2M0335+2342). \cite{2017ApJ...845...33G} used a maximum likelihood estimation (MLE)  to obtain $\alpha$ = 1.6 $\pm$0.2 and  1.8$\pm$0.2 for W1906+40 and 2M0335+2342 respectively. 
Assuming that all the flares with energy in the range from \textit{E$_{min}$}-\textit{E$_{max}$} follow a FFD with a uniform power law, the total energy of all flares during observation time \textit{T} can be computed by using the spectral index $\beta$. This total energy is expressed as \citep{1983Ap&SS..95..235G}:
\begin{equation} \label{energy_final}
\varepsilon = T \times 10^{\alpha}\beta(E_{max}^{1-\beta} - E_{min}^{1-\beta})/(1 - \beta)
\end{equation}
\\
In this paper, we report in  Section \ref{sec:data reduction and analysis} on \textit{Kepler} photometry of our 10 UCD targets, and we use the photometric data to estimate the energies of each flare. In Section \ref{sec:false flare injection}, we discuss artificial flare injection and recovery. In Section \ref{sec:flare rates}, we present estimates of flare rates. In Section \ref{sec:huge flares}, we concentrate on the detailed properties of two superflares in our sample. Discussion of our results is presented in Section \ref{sec:discussion}.
\\
\section{Data reduction and analysis} \label{sec:data reduction and analysis}
\subsection{Targets}\label{sec:Targets}
There are 10 UCDs in our sample. In Table \ref{table:target_char} we list, for each target the full name, the \textit{Kepler} ID (EPIC), the \textit{Kepler} magnitude, the 2MASS \textit{J} magnitude, the \textit{K2} campaign number in which the target was observed, the tangential velocity, the optical spectral type, and the distance to the star. To calculate the tangential velocity for each target, we used the relation \textit{V$_{tan}$} = 4.74$d\mu$, where $d$ (in parsecs) is the distance of each target as given in Table \ref{table:target_char} and $\mu$ is the proper motion. For all but one star, we used proper motions from \cite{2015ApJ...798...73G}: the exception is 2M0326+1919, for which we used \cite{2016ApJ...817..112S}. Our sample contains two M6, three M7, three M8 and two L0 dwarfs. Most of the targets are old low mass stars and some may be brown dwarfs. The distances of two targets 2M2228-1325 and TRAPPIST-1 are taken from the literature. For the remaining stars, distances are estimated using either the \textit{M$_{J}$}/ST (ST = spectral type) relationship from \cite{2012ApJS..201...19D} or \textit{i-z/M$_{i}$} relationship from Schmidt et al. 2018 (in prep). The \textit{i-z/M$_{i}$} relation from Schmidt et al. (2018, in prep) is an updated version of \cite{2010AJ....139.1808S}. It is based on a linear fit to the color magnitude diagram of 64 M5-L8 dwarfs and has typical uncertainties of $\sim$12\%. The ninth column in Table \ref{table:target_char} gives the references we used to identify spectral types of each target. The tenth column indicates whether the distances were obtained from the literature or were estimated using photometry: the reader is referred to the Notes to the Table for explanations of \enquote{mjst} and
\enquote{miiz} in Column 10.
{\normalsize
\begin{table*}
	\caption{Target Properties}
      \begin{tabular*}{\textwidth}{cccccccccc}
     \hline
     \hline
      Name & EPIC & $\tilde{K_{p}}$ & \textit{J} & Cam. \# & \textit{V$_{tan}$} (kms$^{-1}$)$^{a}$ & Spt.  & distance (pc) & Ref. & Remarks\\
       \hline
       	2MASS J22285440-1325178 & 206050032 & 14.66  & 10.77 & 3 & 59 & M6.5 & 11.26 $\pm$ 0.62  & 1 & * \\ 
       	(LHS 523, GJ 4281,  LP 760-3) & & & & & & & & \\
       	 2MASS J22021125-1109461 & 206135809 & 16.72  & 12.36 & 3 & 25 & M6.5 & 22.57 $\pm$4.16  & 2 & mjst \\
       	  2MASS J08352366+1029318 & 211332457& 17.55  & 13.14 &  5 & 23 & M7 & 32.26 $\pm$ 5.95  & 3 & mjst \\
       	  2MASS J22145070-1319590	 & 206053352 & 17.74 & 13.46 & 3& 55  & M7.5 & 33.93 $\pm$ 6.26 & 4 & mjst  \\
       	  2MASS J13322442-0441126 & 212826600 & 16.93 & 12.37 & 6  & 5.0 & M7.5 & 20.54 $\pm$ 3.79 & 2 & mjst  \\
       	  2MASS J23062928-0502285 & 200164267 & 15.91 & 11.40 & 12 & 61 & M8 & 12.10 $\pm$ 0.40& 5 & **\\
       	  (TRAPPIST-1)  & & & & & & & & \\
       	  2MASS J12215066-0843197 & 228754562 & 17.93 &13.52 & 10 & 29  & M8 &  32.11 $\pm$ 5.93 &  4 & mjst  \\
       	  2MASS J03264453+1919309 & 210764183 & 18.08 & 13.12 & 4 & 61 & M8.5 & 24.68 $\pm$ 4.55 & 6 & mjst   \\
       	  2MASS J12212770+0257198& 201658777 & 18.40 & 13.17 & 10  & 14 & L0 & 19.66 $\pm$ 2.67 &  6 & miiz \\
       	  2MASS J12321827-0951502 & 228730045 & 18.85  & 13.73 & 10 & 28 & L0 & 26.41 $\pm$ 4.87 & 6 & mjst  \\
       	 \hline
       	\end{tabular*}
       	\\\\
       	  {\textbf{References}}: (1) \cite{1986ApJ...305..784G}; (2) \cite{2003AJ....126.2421C};  (3) \cite{2002ApJ...564..421B}; (4) \cite{2009AJ....137....1F}; (5) \cite{2016Natur.533..221G}; (6) \cite{2008AJ....136.1290R}
       	  \\
       	 Note: mjst = \enquote{Distance estimated using a combination of 2MASS $J$ and spectral type based on the $M_J$/ST relationship from \cite{2012ApJS..201...19D}.} \\
       	 miiz = \enquote{Distance estimated using a combination $i-z$ color based on the $i-z$/$M_i$ relationship from Schmidt et al. (2018, in prep.).}\\
       	 *Distance taken from \cite{2004AJ....128.2460H}\\
       	 **Distance  taken from \cite{2016Natur.533..221G}\\
       	$^{a}$\textit{V$_{tan}$} was calculated using the relation \textit{V$_{tan}$} = 4.74$d\mu$

\end{table*}
\label{table:target_char}
}
\subsection{\textit{K2} photometry}
All the 10 targets listed in Table \ref{table:target_char} were observed by \textit{Kepler K2} in various campaigns (see the campaign number in Table \ref{table:target_char}) in both long cadence mode ($\sim$30 minute, \cite{2010ApJ...713L.120J}) and short cadence mode ($\sim$1 minute, \cite{2010ApJ...713L.160G}). We used short cadence data to study WLFs on all of our targets. We used a method similar to that described in \cite{2017ApJ...838...22G,2017ApJ...845...33G} to measure the photometry of our targets. In order to estimate \textit{Kepler} magnitude which represents brightness of our targets better than the original \textit{Kepler} magnitude \textit{K$_{p}$} provided in the \textit{Kepler} Input Catalog (KIC), we used the relation $\tilde{K_{p}}$ $\equiv$ 25.3 - 2.5log(flux)  \citep{2015ApJ...806...30L}.  Here, flux is the count rate measured through a 3-pixel radius aperture. $\tilde{K_{p}}$ $\approx$ \textit{K$_{p}$ } for most brighter (e.g., AFGK-type) stars \citep{2017ApJ...838...22G}. \\
\\
Our experience in previous works \citep{2017ApJ...838...22G,2017ApJ...845...33G} show that the standard light curves based on default apertures do not give the best results for the ultracool targets. So we used the target pixel files (TPFs) of each targets available in Mikulski Archive for Space Telescopes (MAST) archive instead of using the standard light curves. We began by estimating the best position of each target in each image frame. We inspected some frames by eye to estimate a threshold value of counts for the target pixels in each frame, and used the astropy-affiliated package \enquote{photutils.daofind} to estimate the centroid position in each frame. We used the median of centroids obtained for all the frames as the best position of our targets in their TPFs. We corrected the offset of centroid position in each frame due to spacecraft motion by using the information recorded as POS$\_$CORR1 and POS$\_$CORR2 in each TPFs. After this, we used another astropy-affiliated photometry package \enquote{photutils.aperture$\_$photometry} to measure the photometry of each target using 2 pixel radius aperture. The same number of pixels were used by \cite{2017ApJ...838...22G,2017ApJ...845...33G} to measure photometry of \textbf{UCDs}. We used only good quality (Quality=0) data points. The median count rate through both the 2-pixel radius aperture (CR2) and the 3-pixel radius aperture (CR3) for each target is given in Table \ref{table:median cnt rates}. CR2 is used for flare analysis of all targets in this paper and CR3 is only used for estimation of $\tilde{K_{p}}$.
\\
\\
\begin{table} 
	\centering
	\caption{Median flux of targets}
	\begin{tabular}{ccc}
	\hline
	\hline
	Name & Median flux (cnts/s) & Median flux (cnts/s)\\
	     		& r = 2 & r = 3 \\
	\hline
	2M2228-1325 &  16463 & 18063\\
	2M2202-1109 & 2561 & 2695\\
	2M0835+1029 &  891 & 1254 \\ 
	 2M2214-1319 &  966 & 1058 \\
	 2M1332-0441  &  1956 & 2220 \\
	 TRAPPIST-1 & 5515 & 5717  \\
	 2M1221-0843 & 809 & 884\\
	 2M0326+1919 & 761 & 771 \\
	 2M1221+0257 & 508 & 577\\
	 2M1232-0951 &  358 & 381\\
	 \hline
	\end{tabular}
\end{table}
\label{table:median cnt rates}
\subsection{Flare detection}
Flare detection in the light curve of targets was a multi-step process. The initial step was to remove any periodic features in the light curve, which might be due to systematic or astrophysical variability. These features add complexity to the light curve, and alter the morphology and duration of flares. We began by smoothing the original light curve of each target by using the Python package \enquote{pandas.rolling$\_$median} \citep{mckinney-proc-scipy-2010} to remove the long term trends mainly caused by systematic errors in the light curve \citep{2016ApJ...829...23D}. We used a window of \textit{w} = 3 d data points \citep{2014MNRAS.445.2698H}.  We fitted this smoothed light curve with a third-order polynomial (as suggested by \cite{2016ApJ...829...23D}) and subtracted from original light curve. We then followed similar method described in \cite{2012ApJ...754....4O} to identify the flares in the smoothed light curve. We calculated relative flux \textit{F$_{rel,i}$} for each data point in the smoothed light curve, defined as:

\begin{equation}
F_{rel,i} = \frac{F_i - F_{mean}}{F_{mean}} 
\end{equation}
where \textit{F$_{i}$} is the flux in \textit{i}th epoch and F$_{mean}$ is the mean flux of the entire light curve of each target. This relative flux was used to identify the flare candidates. We used Lomb-Scargle periodogram to examine any other periodic features which we expect to be mainly due to astrophysical variability, e.g., due to presence of starspots. If any periodic feature was detected, we fitted the smoothed light curve with a sinusoidal function using the dominant period, and subtracted from the smoothed light curve. In this way, we prepared the detrended light curve for our targets. We then calculated a statistic $\phi_{ij}$ for each consecutive observation epoch  (\textit{i,j}) as:
\begin{equation}
\phi_{ij} = \Big(\frac{F_{rel,i}}{\sigma_{i}} \Big) \times \Big(\frac{F_{rel,j}}{\sigma_{j}}\Big) , j = i+1
\end{equation}
Here \textit{$\sigma_{i}$} is the error in the flux which is associated with \textit{i}th epoch. This statistic defined in \cite{1993AJ....105.1813W} and \cite{1996PASP..108..851S}  was used to study variable stars using automated searches. It was later used by \cite{2009AJ....138..633K} and \cite{2012ApJ...754....4O} for flare search in different stars. In order to identify the possible flare candidates in the light curve, we used the false discovery rate (FDR) analysis described in \cite{2001AJ....122.3492M} . This method uses a critical threshold value of the \textit{$\phi_{ij}$} statistic which is different for each target. To calculate this critical value of \textit{$\phi_{ij}$}, we first discarded all those epoch pairs for which \textit{$\phi_{ij}$} $>$ 0 but \textit{F$_{rel,i,j}$} $<$ 0. We then divided the remaining  \textit{$\phi_{ij}$} distribution in two distributions: the null distribution for which \textit{$\phi_{ij}$} $<$ 0 and possible flare candidate distribution for which \textit{$\phi_{ij}$} $>$ 0. The absolute value of the null distribution was fitted by a Gaussian function. The parameters of this Gaussian function was then used to calculate the \textit{p}-values of each \textit{$\phi_{ij}$} in flare candidate distribution.  We then followed each step described in Appendix B of \cite{2001AJ....122.3492M} to calculate the critical \textit{p}-value and hence the critical \textit{$\phi_{ij}$}. Epochs with \textit{$\phi_{ij}$} greater than this value of the critical \textit{$\phi_{ij}$} were considered to be better flare candidates in the light curve. The value of variable $\alpha$ which is used in \cite{2001AJ....122.3492M} FDR analysis is chosen to be 0.05 for \textit{Kepler} data  (based on private communication from R. Osten). This value of $\alpha$ signifies that no more than 5\% of the epochs with \textit{$\phi_{ij}$} greater than the critical \textit{$\phi_{ij}$} are false positives (mostly due to noise in the data). We also used additional criteria, namely, that the detrended flux should exceed the photospheric level by 2.5\textit{$\sigma$}. This decreased the number of flare candidates in our data set to a few hundred. The final flares were chosen by inspecting the data by eye. In this way, even for the weakest flares we ensured that there is at least a pair of epochs for which \textit{F$_{rel,i,j}$} $>$ 0: by this means, we were able to exclude any flares which had only a single measurement of flux brightening. For strong flares, there are multiple consecutive epoches with \textit{F$_{rel,i}$} $>$ 0. More detailed explanation regarding this method of flare detection can be found in \cite{2012ApJ...754....4O}.
\\
\subsection{Calibration of Equivalent Duration and Calculation of Flare Energy}
To calculate the flare energies, we first estimated equivalent duration (hereafter ED) of each flare. It depends on the filter used but is independent of the distance to the flaring object and so it is widely used for determining flare energies. ED of a flare is expressed as:
\begin{equation}
ED = \int{[(F_{f} - F_{c})/F_{c}]dt}
\end{equation}
where \textit{F$_{f}$} is the flare flux and \textit{F$_{c}$} is the continuum flux (i.e when the star is in its quiescent state). It has units of time and gives the area under the flare light curve. It is the equivalent time during which the star (in its quiescent state) would have emitted the same amount of energy as the flare actually emitted \citep{1972Ap&SS..19...75G}. We follow the method described in \cite{2017ApJ...845...33G,2017ApJ...838...22G} to calibrate ED in terms of energy. \textit{Kepler} measures photometry in the wavelength range 430nm to 900nm. In case of UCDs which have lower effective temperatures, a significant part of the flux is contributed by the longer wavelength part of this range but the WLF radiation contributes flux throughout the whole range of wavelengths in the \textit{Kepler} band. This means that a given number of WLF counts measured in the \textit{Kepler} band will have higher mean energy than the same number of  photospheric counts from the UCD \citep{2013ApJ...779..172G}.
For each target, we estimate the photospheric spectrum by using the matching late-M or L dwarf template spectrum 
\citep{2007AJ....133..531B,2014PASP..126..642S} normalized to match the Pan-STARRS $i$-band photometry \citep{2012ApJ...750...99T,2016arXiv161205560C,2016arXiv161205242M}
\footnote{For LHS 523, which is too bright for Pan-STARRS, we normalize to the DENIS $I$-band photometry \citep{1997Msngr..87...27E}}.  
We compute the photospheric specific flux of a 10,000 K blackbody which is normalized to have the same count rate through the \textit{Kepler} filter as the photosphere of each target. (Using an 8,000 K blackbody gives values only 2\% lower: this reduction is much less than other sources of uncertainty.) 
We multiply this photospheric specific flux by the full width half maximum (FWHM) of the \textit{Kepler} band pass (4000$\AA$), 4\textit{$\pi$d$^{2}$} (\textit{d} = distance of target), and the \textit{Kepler} equivalent duration to obtain \textit{Kepler} flare energy \textit{E$_{Kp}$}. Since a 10,000 K blackbody is more energetic for same count rate, we applied a correction factor of 1.3 to get the final estimate of \textit{E$_{Kp}$}. The photospheric specific flux of a 10,000 K blackbody corresponding to each target is given in Table \ref{table:flare stats}. Figure \ref{fig:10Kflare_trappist1_scaling} shows the optical and near-infrared spectral energy distribution of TRAPPIST-1 and a 10,000 K flare with the same count rate through the \textit{Kepler} filter\footnote{For TRAPPIST-1 we used an average distance of 12.3 pc which is the average of distances mentioned in \cite{2016Natur.533..221G} and \cite{2016AJ....152...24W}}.
Computation of total flare energies integrated over the ultraviolet, visible and infrared is useful to compare the results with those obtained by using other surverys. For the 10,000 K blackbody flare model we adopt, this bolometric energy (UV/optical/infrared) is 3.1 times the $E_{Kp}$ we report. Likewise, for an 8,000 K blackbody the factor is 2.5.  \cite{2013ApJ...779..172G} argued this range is simlar to that seen in an M dwarf flare by \cite{1991ApJ...378..725H}. In Table \ref{table:equivalent_duration}, we list the time at which peak flare emission occured, the equivalent duration, and the \textit{Kepler} energy for all of the flares which we identified in our targets. \\
\begin{table} 
	\centering
	\caption{Flare Properties}
	\begin{tabular}{cccc}
	\hline
	\hline
	EPIC & Peak Flare Time & ED  & log $E_{Kp}$\\
	     		& (BJD - 2454833) & s & erg \\
	\hline
	206050032 & 2145.4316 & 1.1e+00  & 29.5 \\
	206050032 & 2145.5278 & 6.3e-01 & 29.3\\
	206050032 & 2146.7530 & 2.4e+02 & 31.9\\
	206050032 & 2147.2441 & 3.1e+01 & 31.0\\
	206050032 & 2147.7570 & 1.5e+00 & 29.7\\
	 \hline
	\end{tabular}
	\\
	 Only a portion of this table is presented here to show \\
	  its form and content. A machine-readable version\\
	  of the full table is available.\\
\end{table}
\label{table:equivalent_duration}
\begin{figure}  
\includegraphics[scale=0.55,angle=0]{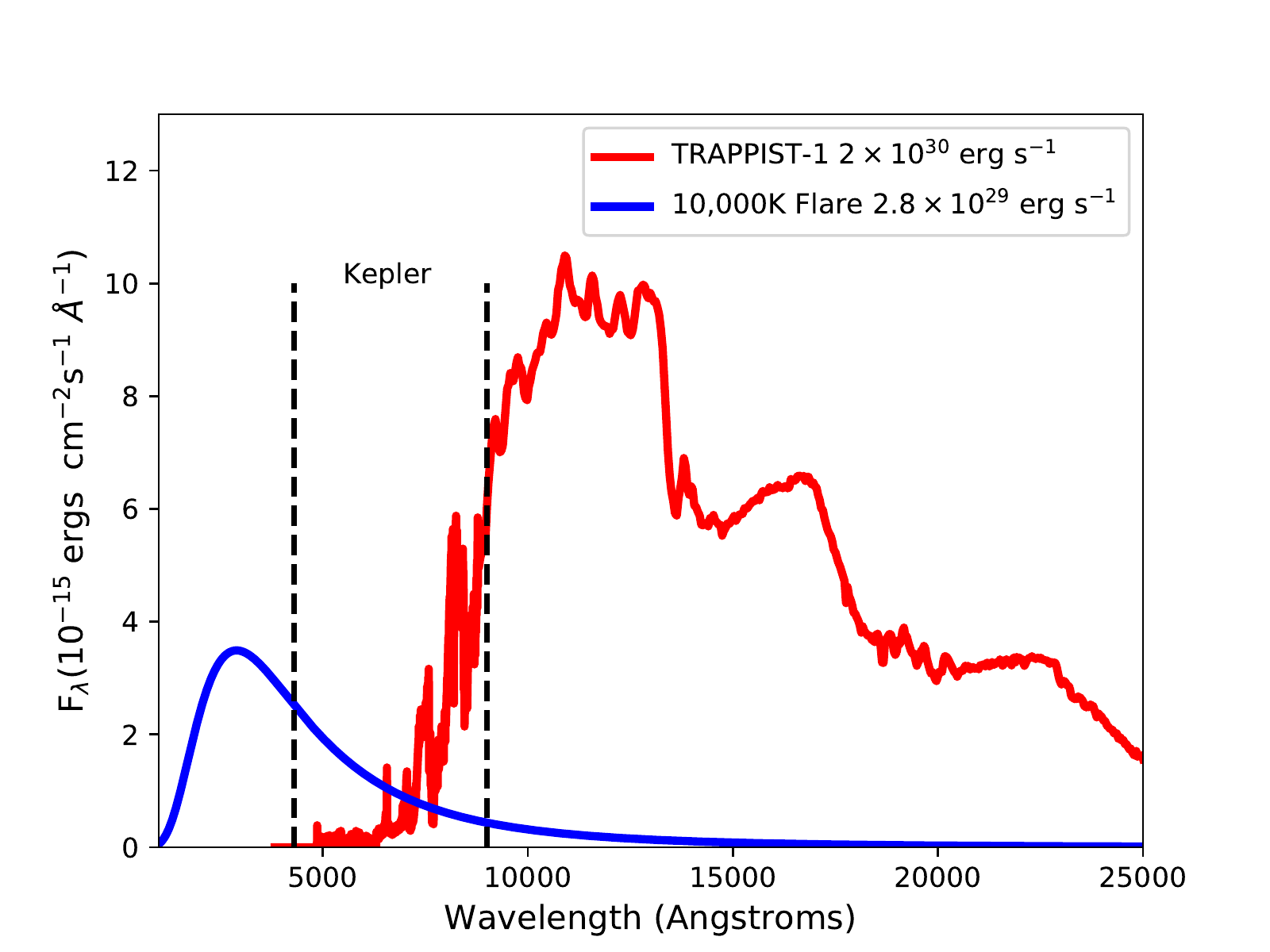}
\caption{The optical and near-infrared spectral energy distribution of TRAPPIST-1 (red) and a hypothetical 10,000 K blackbody (blue). The wavelength range in between vertical dashed lines is \textit{Kepler} band. Using the distance of 12.3 pc, the bolometric luminosity of TRAPPIST-1 is 2.0 $\times$ 10$^{30}$ erg s$^{-1}$ and that of 10,000 K flare is 2.8 $\times$ 10$^{29}$ erg s$^{-1}$.}
\end{figure} 
\label{fig:10Kflare_trappist1_scaling}
\section{Artificial Flare Injection and estimation of lowest detectable flare energy} \label{sec:false flare injection}
In order to get an estimate of the minimum flare energy which could be detected by our algorithm, we generated artificial flares of randomly chosen amplitude and duration using Davenport model \citep{2014ApJ...797..122D}.  This was done by a slight modification of a similar module used in software package known as \enquote{appaloosa} \citep{2016ApJ...829L..31D}. Then we injected the artificial flares at random times to the detrended light curve with 1-$\sigma$ noise. To prepare this detrended light curve, we followed similar detrending process described in Section \ref{sec:data reduction and analysis} and masked all other fluxes greater than 1-$\sigma$ level from the median flux. For simplicity, all injected artificial flares were single-peak \enquote{classic flares}. Care was taken to avoid any overlapping of the injected flares. We injected 10 artificial flares at once and used our algorithm to detect them. We kept track of times at which the artificial flares were injected and their equivalent durations. We repeated this process 1,000 times, so a total of 10,000 artificial flares were generated: however, due to our restriction to non-overlapping events, some fraction of the 10,000 couldn't be injected. We then calculated the flare energies of the injected artificial flares and compared these values with the energies we recovered by means of our algorithm. We found that weak flares having energies less than a certain energy were not detected by our algorithm. The light curves of different targets have different noise levels, so the minimum detectable energy of weak flares as found by our algorithm is different for each target. To estimate this minimum energy, we repeated the above process separately for each target. A list of minimum energies of artificial flares injected and later detected by our algorithm is given in Table \ref{table:false flare min energies}. For flare analysis, we discarded all the flares (if any) having energies less than the minimum energy of the artificial flares detected by our algorithm.
\\
\begin{figure*}
\includegraphics[scale=0.60,angle=0]{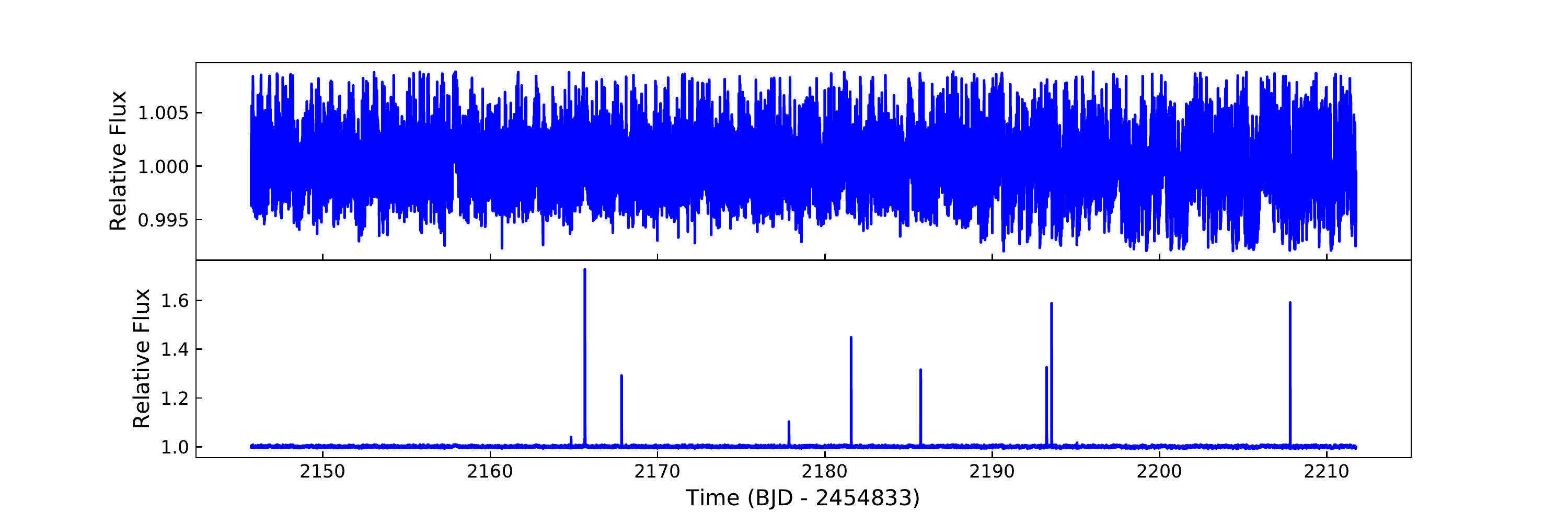} 
\caption{An example of artificial flare injection in 2M2228-1325 light curve. The upper plot is detrended light curve with 1-$\sigma$ noise and the lower plot is the light curve with artificial flares injected at random times.}
\end{figure*}
\label{fig:false_flare_inj}
\begin{table} 
	\centering
	\caption{Minimum energies of injected and detected artificial flares}
	\begin{tabular}{ccc}
	\hline
	\hline
	Name & \textit{E$_{Kp},min$} (10$^{30}$ erg) & \textit{E$_{Kp},min$} (10$^{30}$ erg)\\
	& injected & detected \\
	\hline
	2M2228-1325 & 0.014 & 0.18 \\
	2M2202-1109 & 0.008 & 0.65  \\
	2M0835+1029 & 0.006 & 0.93 \\ 
	 2M2214-1319 & 0.004 & 1.3  \\
	 2M1332-0441  &  0.009 & 0.59  \\
	 TRAPPIST-1 &  0.004 & 0.17 \\
	 2M1221-0843 &  0.002 & 1.3\\
	 2M0326+1919 & 0.021 & 0.62 \\
	 2M1221+0257 & 0.006 & 0.37 \\
	 2M1232-0951 & 0.004 & 0.83 \\
	 \hline
	\end{tabular}
\end{table}
\label{table:false flare min energies}
\section{Flare statistics and Flare energy spectrum of target UCDs} \label{sec:flare rates}
Table \ref{table:flare stats} lists various properties of flares on sample targets. In this table, \enquote*{specific flux} is the flux per cm$^{2}$ per second per Angstrom for a 10,000 K blackbody which has the same number of counts as the target in \textit{Kepler} bandpass. The column \enquote*{Periodic Features} gives information if any periodic feature is seen in the light curve of each target after removing the long term trends using rolling median method. If any periodicity is seen, the dominant period is listed. The periodicity might be due to the presence of starspots, and most probably gives an indication of the rotation period of the target. \textit{N} is the number of flares observed on a given target during the entire interval \textit{T} of \textit{K2} observations of that target (for \textit{T} values, see Table \ref{table:flare parameters}). Likewise, \textit{E$_{Kp,min}$} is the minimum \textit{Kepler} energy of all the identified flares on a given target and \textit{E$_{Kp,max}$} is the maximum \textit{Kepler} energy of all the identified flares on a given target. \\ \\
\begin{table*} 
	\centering
	\caption{Flare statistics}
      \begin{tabular}{cccccc}
     \hline
     \hline
      Target & Sp. flux & Periodic feature & \textit{N} & \textit{E$_{Kp},min$}  & \textit{E$_{Kp},max$}  \\
       & (erg cm$^{-2}$ s$^{-1}$ $\AA^{-1}$) & & & (10$^{30}$ erg) & (10$^{30}$ erg) \\
       \hline
       2M2228-1325 & 3.66 $\times$ 10$^{-15}$ & No & 50 & 0.19 & 72\\
       	2M2202-1109 & 4.64 $\times$ 10$^{-16}$ & Yes (0.42 d) & 50 & 0.64 & 310 \\
       	2M0835+1029 &  1.55 $\times$ 10$^{-16}$  & No & 11 & 1.1 & 2900 \\ 
       	 2M2214-1319 &  1.64 $\times$ 10$^{-16}$  & No & 26 & 2.2 & 350\\
       	 2M1332-0441  & 3.62 $\times$ 10$^{-16}$  & No & 31 & 0.69 & 85  \\
       	 TRAPPIST-1 & 8.95 $\times$ 10$^{-16}$  & Yes (3.3 d)& 39 & 0.21 & 230\\
       	 2M1221-0843 & 1.47 $\times$ 10$^{-16}$ & Yes (0.27 d)& 36 & 1.4 & 150 \\
       	 2M0326+1919 & 1.40 $\times$ 10$^{-16}$ & Yes (0.97 d)& 18 & 0.67 & 55 \\
       	 2M1221+0257 & 1.10 $\times$ 10$^{-16}$ & Yes (0.18 d) &  11 & 0.48 & 25\\
       	 2M1232-0951 &  6.29 $\times$ 10$^{-17}$ & No & 11 & 0.82 & 1200\\
       \hline
       \end{tabular}
 \end{table*}
\label{table:flare stats}
Table \ref{table:flare parameters} lists the values of fitted parameters for the FFD of each of our targets. In this table \textit{T} is the total observation time for a given target, $\beta$ and $\alpha_{o}$ are the fitted values of parameters in Eq. \ref{eq:cu_power_law}. Likewise, \textit{E$_{min}$}  and \textit{E$_{max}$} are the minimum and maximum \textit{Kepler} energies used for fitting. The total observation time \textit{T} (in seconds) is computed by counting the total number of good (Quality = 0) data points and multiplying by 58.85 s which is the correct exposure time equivalent to 1 short cadence. We used the maximum-likelihood method described in \cite{2010arXiv1008.4686H} and implemented in the routine known as \enquote{emcee} \citep{2013PASP..125..306F} to fit a straight line to our data (in log scale) and hence obtain the optimal values of parameters $\beta$ and $\alpha_{o}$. Here we report the intercept $\alpha_{o}$ corresponding to energy 10$^{30}$ erg, not the zero energy. It will be helpful in comparing the flare rates of our targets with previously reported flare rates of other targets observed by \textit{Kepler and K2}, most of which have flare energies greater than 10$^{30}$ erg \citep{2013MNRAS.434.2451R,2014ApJ...797..121H,2016ApJ...829...23D,2017ApJ...845...33G}. The routine \enquote{emcee} uses the standard Metropolis-Hastings Markov-Chain Monte Carlo (MCMC) procedure for marginalization and uncertainty estimation. We neglected the highest observed energy for fitting the line to reduce any bias in the analysis. Since 2M2228-1325 has a heavy-tailed distribution, we need to select a minimum value of energy to be considered for fitting. This minimum value  was chosen on the basis of fitting for broken power law discussed below. For targets which do not have a heavy tailed distribution, all flares, even those with the lowest energies, were considered for fitting. 
We also used the analytic solution to get the estimate of parameter $\alpha$ which is derived in \cite{arXiv:0706.1062} and references therein. Here we denote this estimate as $\hat{\alpha}$ which is the maximum likelihood estimator (MLE) of true parameter $\alpha$ and is expressed as: 
\begin{equation}\label{eq:alpha_hat}
\hat{\alpha} = 1 + n \bigg[\sum_{i=1}^{n}  ln \frac{E_{i}}{E_{min}}\bigg]^{-1}
\end{equation}
with error
\begin{equation}
\sigma = \frac{\sqrt{n+1}(\hat{\alpha} - 1)}{n}
\end{equation}.
Here \textit{n} is the number of flares and $E_{i}$, i = 1...\textit{n} are the observed values of energies \textit{E} such that \textit{$E_{i}$} $\geq$ \textit{$E_{min}$}. The MLE solution $\hat{\alpha}$ is an unbiased estimator of $\alpha$ in the asymptotic limit of large sample size, \textit{n} $\rightarrow$ $\infty$ . In addition, a more reliable estimate for parameter $\alpha$ can be obtained for sample size n $\gtrsim$ 50  \citep{arXiv:0706.1062}. For small sample sizes \textit{n}, \cite{2015stam.book.....A} suggests the value of $\hat{\alpha}$ obtained by using Eq. \ref{eq:alpha_hat} can be multiplied by a factor of $(n-2)/n$ in order to make the result unbiased. The unbiased values of $\hat{\alpha}$ for each target are listed in Table \ref{table:flare parameters}. The same energy range used for fitting the power law FFD is used for estimating $\hat{\alpha}$. As the number of flares on most of our targets is unfortunately rather small for a more accurate estimation of parameter $\alpha$ using the analytic solution: we will use the results obtained by using the \enquote{emcee} routine for discussion and comparison of our results with previous works. Table \ref{table:flare parameters} also lists the slopes of FFDs for 2M0335+2342 and W1906+40, obtained by using the methods described above. It should be noted that they are slightly different from the values reported in previous papers because somewhat different energy intervals were chosen for fitting.\\ \\
\begin{table*} 
	\centering
	\caption{Power law fits to FFDs of targets}
	\begin{tabular}{cccccccc}
	\hline
	\hline
	Name & \textit{ T } & $\beta$ & $\alpha$=$\beta$+1 & $\alpha_{o}$ & $\hat{\alpha}$ & \textit{E$_{min}$}  &\textit{E$_{max}$} \\
	 & (days) & & & & & (10$^{30}$ erg) & (10$^{30}$ erg) \\
	\hline
	2M2228-1325 & 66.8 & 1.03$\pm$0.05 & 2.03 & -1.14$\pm$0.05 & 1.90$\pm$0.18 &3.6 & 32  \\
	2M2202-1109 & 67.5 & 0.65$\pm$0.02 & 1.65 & -1.50$\pm$0.02  & 1.49$\pm$0.07& 0.64 & 100 \\
	2M0835+1029 & 73.3 & 0.65$\pm$0.05  & 1.65 & -2.11$\pm$0.03   & 1.43$\pm$0.14& 1.1 & 15 \\ 
	 2M2214-1319 & 66.9 & 0.70 $\pm$0.02 & 1.70 & -1.45$\pm$0.02  & 1.54$\pm$0.11 & 2.2 & 160 \\
	 2M1332-0441  & 76.4 & 0.57$\pm$0.03 & 1.57 & -1.81$\pm$0.02  & 1.49$\pm$0.09 & 0.69 & 38  \\
	 TRAPPIST-1 &  70.6 & 0.61$\pm$0.02 & 1.61 & -1.94 $\pm$0.01  & 1.47$\pm$0.08 & 0.21 & 18 \\
	 2M1221-0843 & 53.4 & 0.75 $\pm$0.02 & 1.75 & -1.37$\pm$0.02  & 1.66$\pm$0.12 & 1.4 & 49  \\
	 2M0326+1919 & 69.2 & 0.45 $\pm$0.04 & 1.45 & -1.98$\pm$ 0.03  & 1.32$\pm$0.08 & 0.67 & 44 \\
	 2M1221+0257 & 53.5 & 1.04$\pm$0.08 & 2.04 & -2.34$\pm$0.02  & 1.85 $\pm$0.28 & 0.48 & 2.3\\
	 2M1232-0951 & 53.5 & 0.34$\pm$0.04 & 1.34 & -2.16 $\pm$0.03 & 1.37 $\pm$0.12 & 0.82 & 130\\
	 2M0335+2342 & 69.0 & 0.64 $\pm$0.03 & 1.64 & -1.23 $\pm$0.05 & 1.77$\pm$0.19 & 15 & 320 \\
	 W1906+40 & 83.0 & 0.52 $\pm$0.05 & 1.52 & -2.12 $\pm$0.03 & 1.37 $\pm$0.09 & 0.56 & 19 \\
	 \hline
	\end{tabular}
	\\
	Note: The intercept $\alpha_{o}$ reported in this table gives the flare rate for flare energy 10$^{30}$ erg, not the zero energy. See the text for more description.
\end{table*}
\label{table:flare parameters}
Figure \ref{fig:ffds_individual} shows the FFD of each target in the sample. The FFD for each star is presented separately in each panel, and the panels are arranged in order of increasingly late spectral type. The FFDs of other targets in the sample are also plotted in the background of each panel to enable the reader to compare the FFD of a given target with others in our sample. Figure \ref{fig:ffds all} compares FFDs of all the targets in sample. It should be noted that the fitted line of 2M1221+0257 covers a very small range of energy which is less than one order in magnitude. Likewise, Figure \ref{fig:compare M7_L0 FFD} shows a comparison of the FFD for a young brown dwarf 2M0335+2342 with the FFD for an L0 dwarf 2M1232-0951.
\begin{figure*} 
\includegraphics[scale=0.75,angle=0]{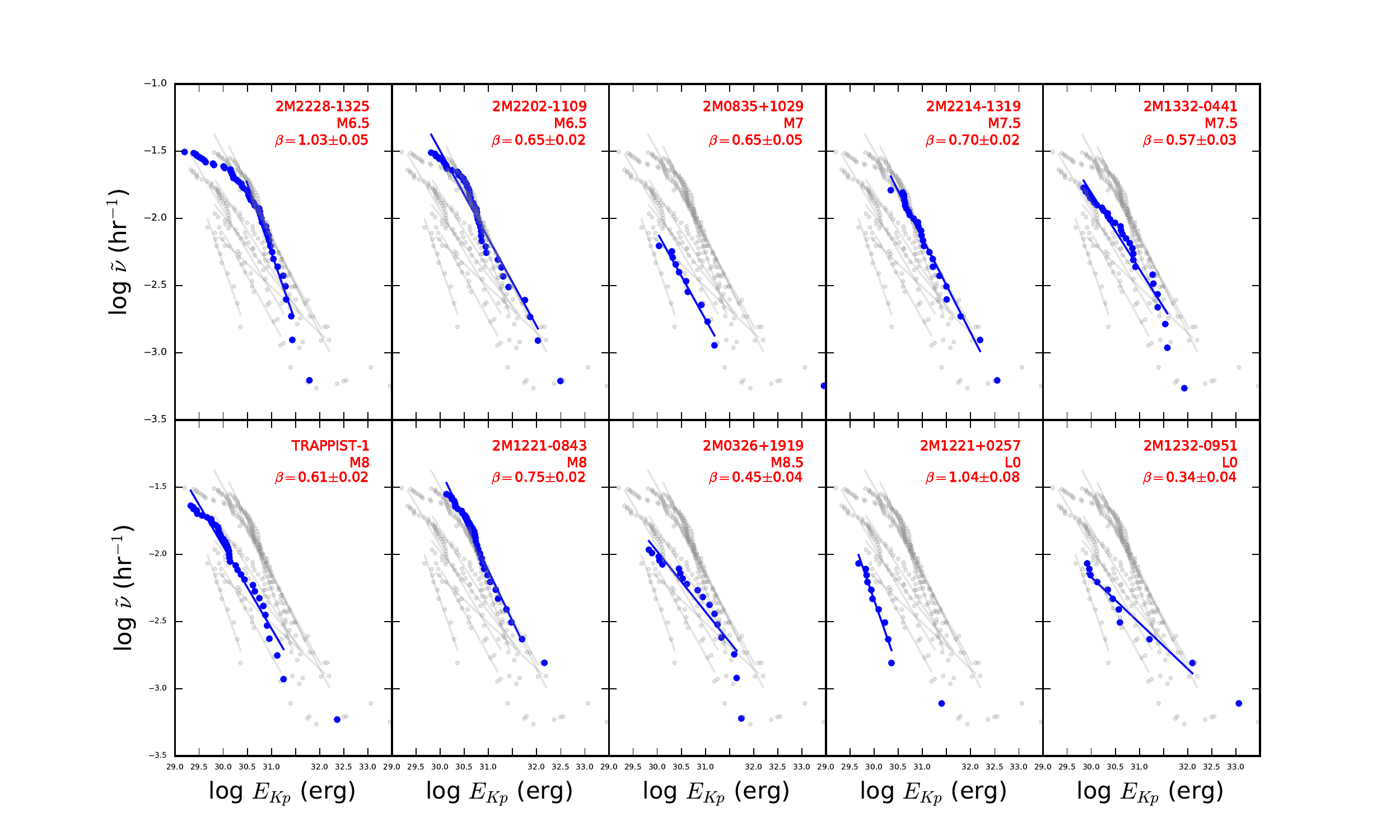} \caption{Individual FFD of all UCDs discussed in this paper. The FFDs are presented according the spectral order of targets. The blue dots represent the observed data and the blue solid line represents the fitted model in each plot. The FFDs of all remaining targets are also plotted in the background to make easy for comparison.}
\end{figure*}
\label{fig:ffds_individual}
%

%
\begin{figure*}
\includegraphics[scale=0.80,angle=0]{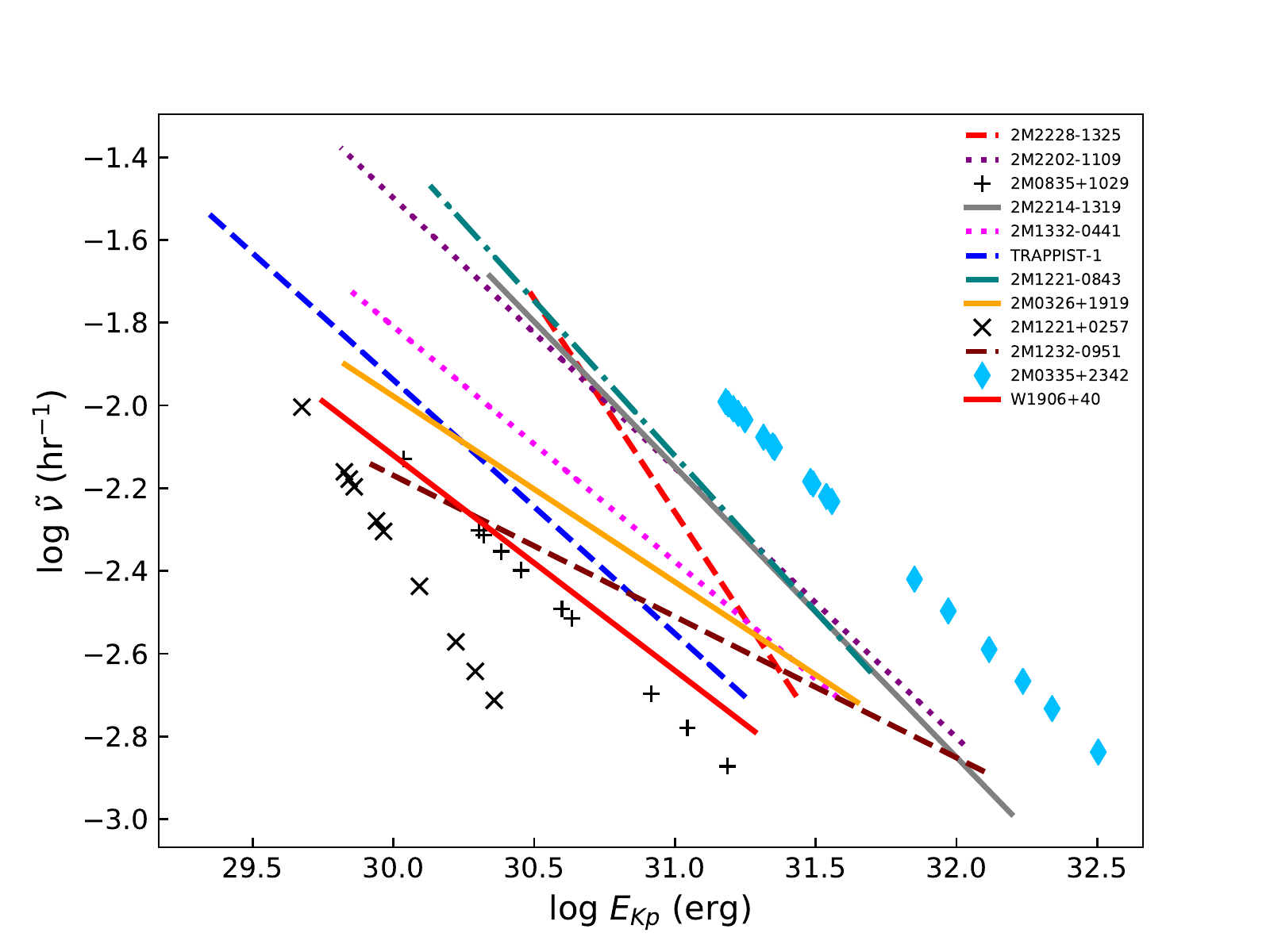} 
\caption{Comparison of fitted FFDs of all UCDs in our sample. The maximum observed energy was not included to minimize biasness. The energy range used for fitting is given in Table \ref{table:flare parameters}. FFDs of two more UCDs: W1906+40 and 2M0335+2342 are also included here. We used the method described in this paper to fit their FFDs.}
\end{figure*}
\label{fig:ffds all}
\begin{figure} 
\includegraphics[scale=0.55,angle=0]{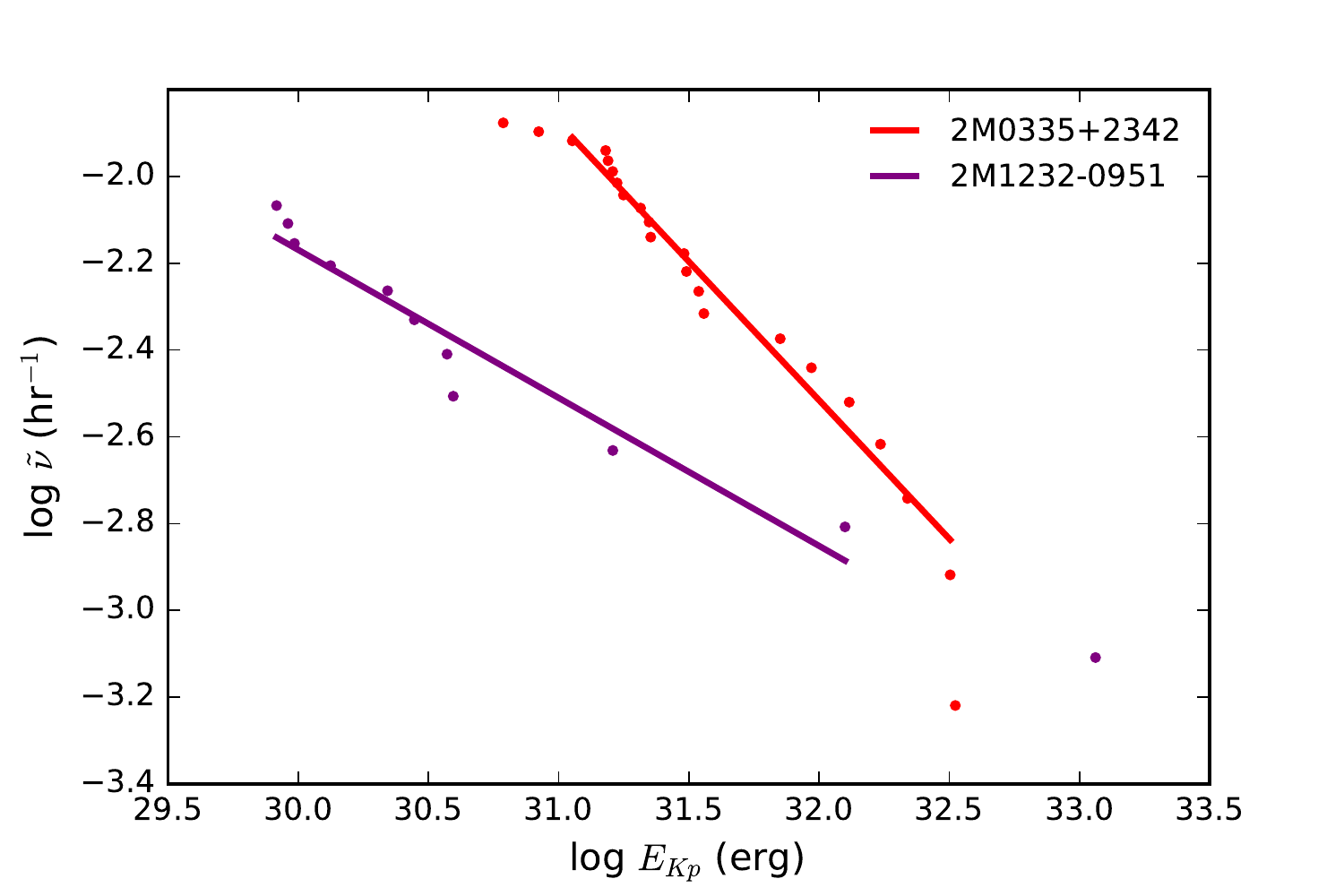} \caption{Comparison of FFD of a young brown dwarf 2M0335+2342 (red) and L0 dwarf 2M1232-0951 (purple). The dots represent observed data and solid lines represent the fitted power law model. This plot suggests that 2M1232-0951 has flare rate comparable to a 24 Myr young brown dwarf in case of high energy flares despite having cool atmosphere.}
\end{figure}
\label{fig:compare M7_L0 FFD}
\subsection{Possibility of a broken power law or a power law with exponential cutoff  in the FFD of some UCDs}
The presence of long tails at high energies in the FFDs suggests that a single power law might not be the optimal fit to the FFD of all UCDs. In the case of 2 of our targets (2M2228-1325 and 2M0326+1919), we tried to fit two different models: (i) a broken power law, and (ii) a power law combined with an exponential cutoff. Results are shown in Figures \ref{fig:broken powerlaw:2M2228-1325} and \ref{fig:broken powerlaw:2M0326+1919}. The broken power law model can be expressed as:
\begin{equation}
	f(E) =
	\begin{cases}
			 A(E/E_{break})^{-\alpha_{1}}  : E < E_{break}  \\
              A(E/E_{break})^{-\alpha_{2}}  : E > E_{break}
	\end{cases}
\end{equation}
while the power law with exponential cutoff can be expressed as:
\begin{equation}
f(E) = A(E/E_{0})^{-\alpha}exp(-E/E_{cutoff})
\end{equation}
We used the astropy-affiliated packages \enquote{modeling.power- laws.BrokenPowerLaw1D} and \enquote{modeling.powerlaws.\\ExponentialCutoffPowerLaw1D} to estimate the model parameters. And we used the astropy-affilicated package \enquote{modeling.fitting.LevMarLSQFitter} to fit the data. The later package uses the Levenberg-Marquardt algorithm and least squares statistic for fitting. We used a likelihood ratio test to decide which model better fits the observed data, by considering the broken power law model as alternate, and the power law with exponential cutoff as the null model. In the case of 2M2228-1325, we found that the broken power law provides a better fit to the data with \textit{$\alpha_{1}$} = 0.20, \textit{$\alpha_{2}$} = 0.83, \textit{E$_{break}$} = 3.25 $\times$ 10$^{30}$ erg and \textit{A} = 29.86. The p-value of the likelihood-ratio test statistic is 0.005 (i.e. we reject our null model for a significance level of 0.05). We used the energies greater than \textit{E$_{break}$} for determining the slope of the FFD. But, in the case of 2M0326+1919, we found that the power law with exponential cutoff provides a better fit to the data. In this case, the fitted parameters are \textit{A} = 15.15, \textit{E$_{cutoff}$} = 3.83 $\times$ 10$^{31}$ erg, \textit{E$_{0}$} = 1.40 $\times$ 10$^{30}$ erg and \textit{$\alpha$} = 0.23. The p-value of the test statistic is 0.60 (i.e. we accept our null model for a significance level of 0.05). The possible break in the power law could be due to a number of factors, including the sensitivity of the instrument in observing weak flares, or saturation, or an upper limit on the energy which any flare on a given target is able to release \citep{2005stam.book.....G}. As far as we have been able to determine, broken power law FFD's laws are seldom discussed in literature. 
\begin{figure}
\includegraphics[scale=0.55,angle=0]{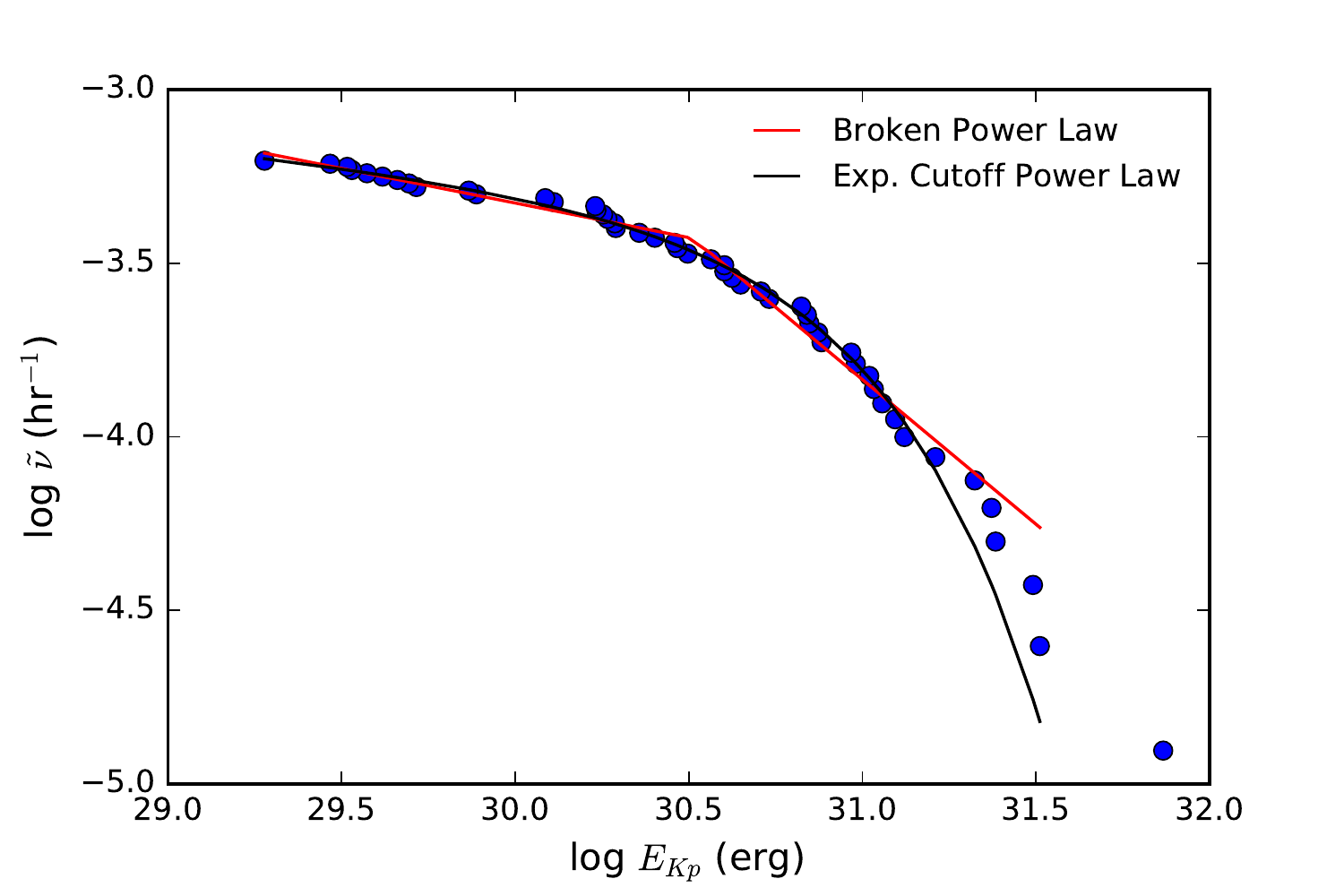} 
\caption{Comparison between broken power law and power law with cutoff in case of 2M2228-1325.}
\end{figure}
\label{fig:broken powerlaw:2M2228-1325}
\begin{figure}
\includegraphics[scale=0.55,angle=0]{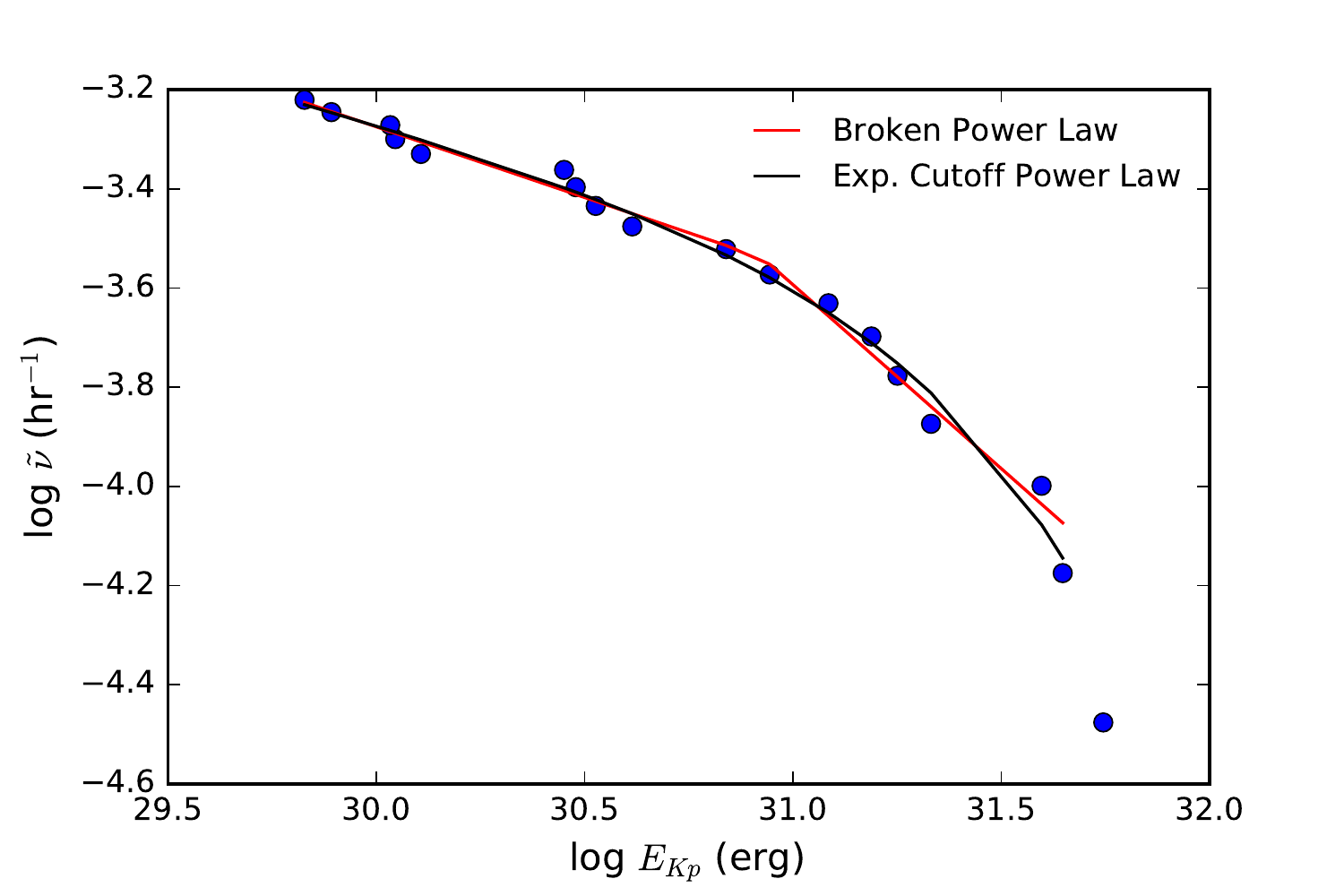} 
\caption{Comparison between broken power law and power law with cutoff in case of 2M0326+1919.}
\end{figure}
\label{fig:broken powerlaw:2M0326+1919}
\section{Superflares observed on two UCDs}\label{sec:huge flares}
One of the advantages of short cadence \textit{Kepler} data is that it helps to study the time scales associated with rapid rise, rapid decay and gradual decay phase of superflares (flares with energies $\geq$10$^{33}$ erg). Here we present the photometry of two superflares observed on our targets. 
\subsection{Photometry of superflare observed on L0 dwarf  2MASS J12321827-0951502}
 We observed a superflare on L0 dwarf 2M1232-0951 at \textit{Kepler} time 2811.7662319, when the flare flux rose to 51529 cnts/s  ($\tilde{K_{p}}$ = 13.44, $\Delta \tilde{K_{p}}$ = -5.41): this count rate is $\sim$144 times larger than the photospheric level (358 cnts/s). To determine the relevant time-scales of the flare, we used \cite{2014ApJ...797..122D} (hereafter D14) model to fit the flare light curve. This yields values for the time scale \textit{t$_{1/2}$} associated with the rise of flux from and return to  half maximum flux. The model also yields values for two decay time scales, one rapid (close to flare maximum), and the other gradual (later in the flare). The D14 model uses a flare template that was based on the flare properties of M4 dwarf GJ 1243 using \textit{Kepler} short cadence data. In this model, the rapid rise phase is best fitted by using a fourth order polynomial, and the decay phase is best fitted by using sum of two exponentials  which can be expressed as:
\begin{equation} \label{eq:Davenport14 model}
\Delta F = A(\alpha_{i}e^{-\gamma_{i}\Delta t/t_{1/2}} + \alpha_{g}e^{-\gamma_{g}\Delta t/t_{1/2}} )
\end{equation} 
where \textit{$\Delta$F} is flare only flux, \textit{A} is flare amplitude and \textit{$\Delta$t = t - t$_{f}$} (\textit{t$_{f}$} is the peak flare time). In D14 template, the values of the different parameters of Eq.(\ref{eq:Davenport14 model}) are \textit{$\alpha_{i}$} = 0.6890($\pm$0.0008), \textit{$\alpha_{g}$} = 0.3030($\pm$0.0009), \textit{$\gamma_{i}$} = 1.600($\pm$0.003) and \textit{$\gamma_{g}$} = 0.2783($\pm$0.0007).
\cite{2016ApJ...828L..22S} used this model to estimate the value of \textit{t$_{1/2}$} to be in the range 3 (best fit) to 6.2 (minimal fit)  minutes for a superflare observed on L0 dwarf ASASSN-16ae. Likewise, \cite{2017ApJ...838...22G}  estimated \textit{t$_{1/2}$} = 6.9 minutes for L1 dwarf W1906+40 using same model. Slightly different value of parameters were used by \cite{2017ApJ...838...22G} to fit a superflare on the L1 dwarf SDSSp J005406.55-003101.8 to estimate \textit{t$_{1/2}$} = 7.8 minutes. Here, the estimation of \textit{t$_{1/2}$} for 2M1232-0951 is done using \textit{Kepler} short cadence data, so it is more accurate than those reported for ASASSN-16ae and SDSSp J005406.55-003101.8. To fit the flare of interest to us here (i.e. the superflare on 2M1232-0951), we started by examining if the D14 model could fit the observed data. However, we found that it did not provide an especially good fit for the late decay phase. To get an initial estimation of decay phase time scale, we fitted the late decay phase separately by a single exponential curve and used the parameters obtained in this way to fit the entire curve.
Figure \ref{fig:big flare 2M1232-0951} shows the observed flux and fitted model\footnote{This curve is based on medians of parameters of at least 1.5 million samples generated by using Markov Chain Monte Carlo sampling of the posterior function using emcee (\cite{2013PASP..125..306F} with uniform priors). All the flare fits presented in this paper are fitted using this method. For better results, we ensured that the mean acceptance fraction of the sample ensemble to be between 0.25 and 0.5 as mentioned in emcee documentation.}. At first glance in Figure \ref{fig:big flare 2M1232-0951}(a), there seems to be a good agreement between observed data and fitted model. But, the discrepancy can be clearly seen in log-log version of same plot as shown in Figure \ref{fig:big flare 2M1232-0951}(b). The values of fitted parameters for this flare are: \textit{$\alpha_{i}$} = 0.9691, \textit{$\alpha_{g}$} =  0.0310, \textit{$\gamma_{i}$} = 0.4551 and \textit{$\gamma_{g}$} = 0.01543. Likewise other fitted parameters are \textit{A}= 51005.76, \textit{t$_{1/2}$} = 1.054 minute and time of flare = 2811.7661 days. The observation shows that the flux decreases from its maximum value 51529 cnts/s to about one-half of its maximum value which is 27370 cnts/s in an interval of about 1 minute. Since the best cadence time for gathering \textit{K2} data is also about 1 minute, we estimate that a more accurate value of \textit{t$_{1/2}$} could be around 2 minutes for this superflare. The ED of this flare is 11.4 hours and its total bolometric (UV/optical/infrared) energy is 3.6 $\times$ 10$^{33}$ erg. The total flare duration is 4.2 hours.\\
Using the above fitted parameters and solving for the value of \textit{$\Delta$t = t-t$_{f}$} at which impulsive decay switches to gradual decay, we get \textit{$\Delta$t} = 7.83\textit{t$_{1/2}$} after the peak flare which occurs at  \textit{$\Delta$t} = 0. Now using this time reference, we found that the rise phase contains 22.15\%, the impulsive decay phase 38.70\% and the gradual decay phase 39.15\% of the total energy.
\begin{figure*} 
\centering
\includegraphics[scale=0.65,angle=0]{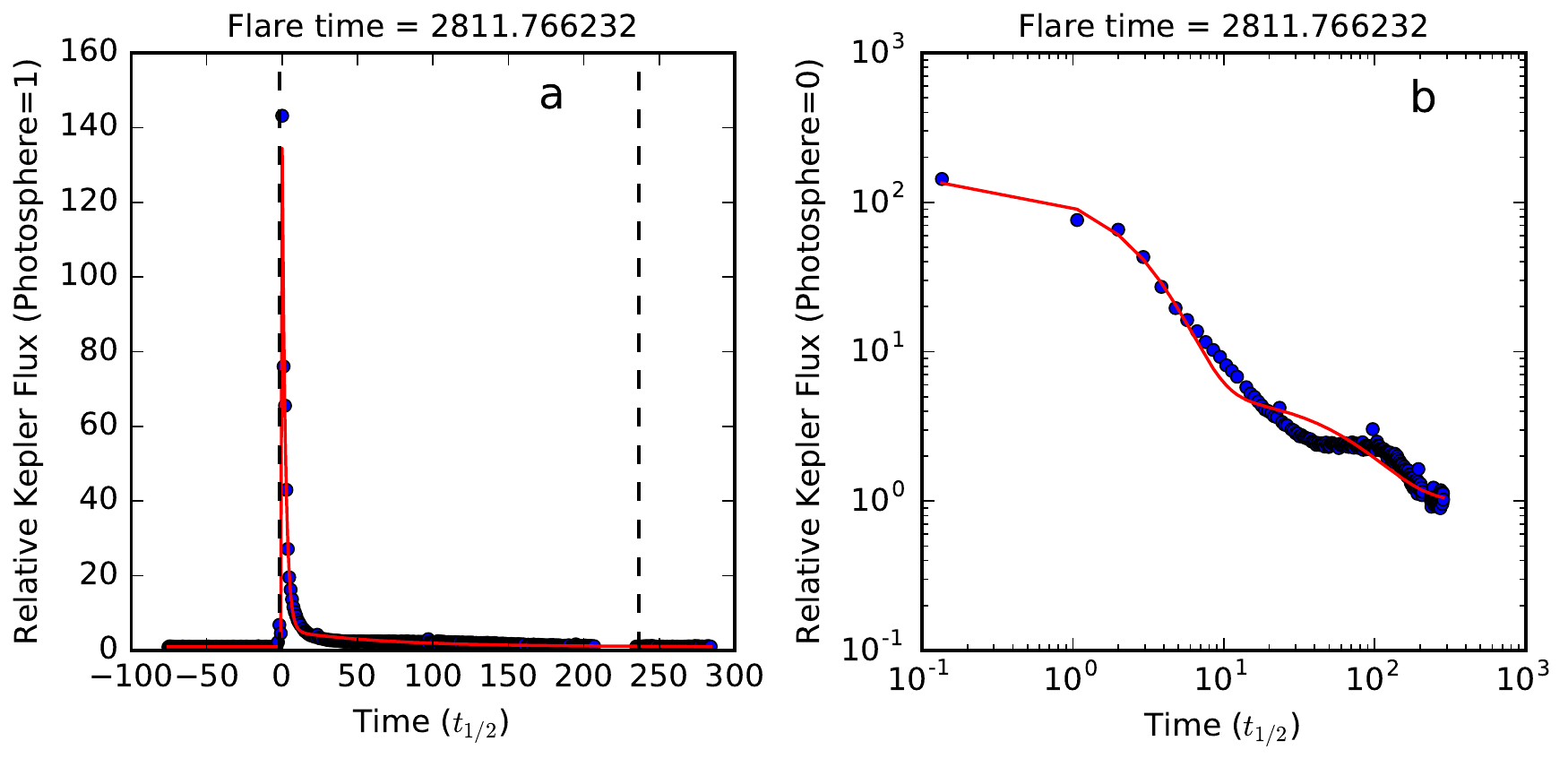} 
\caption{(a) Superflare observed on 2M1232-0951. The blue dots represent the observed flux and the red curve represents the fitted flux using slightly different parameters in D14 model. The time is zero centered at peak flare time and scaled by \textit{t$_{1/2}$}. The vertical dashed lines represent the flare start and end times. (b) This plot is log-log version of plot shown in (a).}
\end{figure*}
\label{fig:big flare 2M1232-0951}
%
%
\subsection{Photometry of superflare observed on M7 dwarf  2MASS J08352366+1029318}
We also observed a superflare on M7 dwarf 2M0835+1029 at \textit{Kepler} time 2379.88288303, when the flare flux is 53213 cnts/s ($\tilde{K_{p}}$ = 13.26, $\Delta \tilde{K_{p}}$ = -4.29): this peak flux is $\sim$60 times larger than the photospheric level (891 cnts/s). For this flare also, we found the original parameters of D14 model do not fit the data particularly well. We consider that we need to make slight changes to the D14 values. We used a fitting procedure similar to that used for the 2M1232-0951 superflare discussed in Section 5.1. We found the following values for the fitted parameters of the M7 superflare: \textit{$\alpha_{i}$} = 0.8182, \textit{$\alpha_{g}$} = 0.1818, \textit{$\gamma_{i}$} = 0.6204 and \textit{$\gamma_{g}$} = 0.08467. Likewise, we found the other fitted parameters are \textit{A}= 58480, \textit{t$_{1/2}$} = 1.73 minutes and time of flare = 2379.8825 days. Both the observed flux and fitted model for this superflare are shown in Figure \ref{fig:M7 dwarf huge flare}(a). This figure and its corresponding log-log version Figure \ref{fig:M7 dwarf huge flare}(b) show that there is better agreement between the observed flare curve and fitted model than in case of the superflare observed on the L0 dwarf 2M1232-0951. The ED of this flare is 7.7 hours and its total bolometric energy (UV/optical/infrared) is 8.9 $\times$ 10$^{33}$ erg. The total flare duration is 4.5 hours. \\
In this superflare, we found that that the value of \textit{$\Delta$t = t-t$_{f}$} at which transition between impulsive decay and gradual decay takes place  is  \textit{$\Delta$t} = 2.81\textit{t$_{1/2}$} after the peak flare (\textit{$\Delta$t} = 0). Now using this time reference, we found that the rise phase contains 19.57\%, the impulsive decay phase contains 25.86\%  and the gradual decay phase 54.58\% of the total flare energy.
\begin{figure*} 
\centering
\includegraphics[scale=0.65,angle=0]{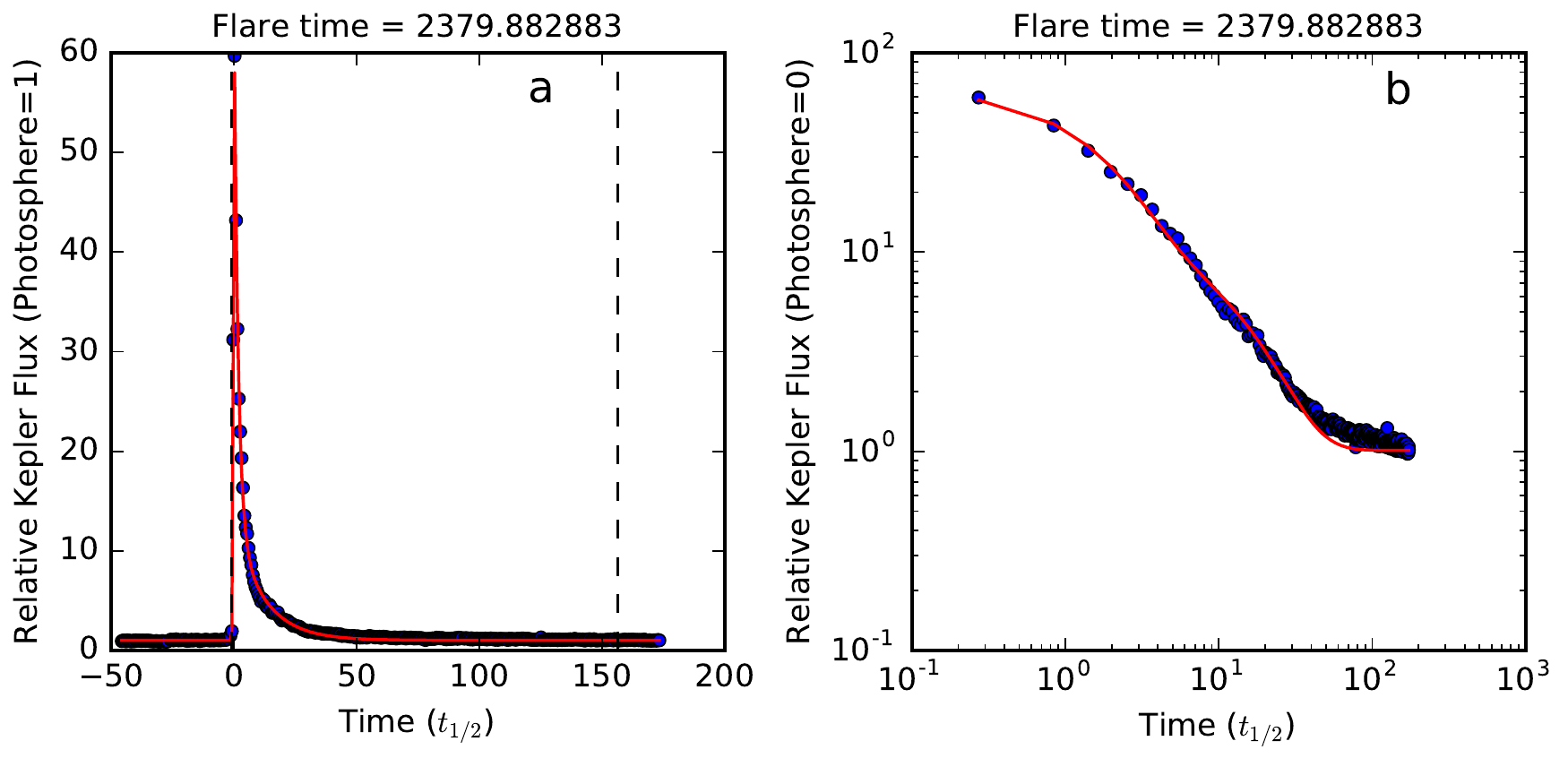} 
\caption{(a) Superflare observed on 2M0835+1029. The blue dots represent the observed flux and the red curve represents the fitted flux using slightly different parameters in D14 model. The vertical dashed lines represent the start and stop times of flare. The time is zero centered at peak flare time and scaled by \textit{t$_{1/2}$}. (b) This plot is log-log version of plot shown in (a).}
\end{figure*}
\label{fig:M7 dwarf huge flare}
%
%
\section{Discussion and Conclusions} \label{sec:discussion}
We identified a total of 283 white light flares in our sample of 10 UCDs. The flares we detected have \textit{Kepler} energies in the range log \textit{E$_{Kp}$} $\sim$(29 - 33.5) erg, and the flares follow power law distributions with slopes $-\alpha$ lying in the range -1.3 \& -2.0. The values of these slopes were determined by using maximum-likelihood method of fitting a straight line, as implemented in \enquote{emcee} software. We also estimated the values of slopes using an analytical method. The presence of noise in the data makes it difficult to detect the weakest flares in UCDs. Due to this, the minimum detectable energy is $>$10$^{29}$ erg in our targets. The late-M dwarfs in our sample have FFD comparable to active mid and late-M dwarfs studied by \cite{2011PhDT.......144H}. Compared to \cite{2011PhDT.......144H}, the flares observed by \textit{Kepler} have higher energy and are less frequent. Yet, we find similar slopes (-$\alpha$). The slopes of late-M dwarfs lie in between -1.5 \& -1.8 except for M6.5 dwarf 2M2228-1325 for which -$\alpha$ = -2.0. We should note that the slopes also depend on the range of energy chosen for fitting. We also analyzed flares on TRAPPIST-1 using the official \textit{K2} pipeline reduced data. We identified 39 good flares on it with bolometric (UV/optical/infrared) flare energies in the range from 6.5 $\times$ 10$^{29}$ to 7.2 $\times$ 10$^{32}$ erg. We find that its FFD has a slope of -$\alpha$ = -1.6. Previously, \cite{2017ApJ...841..124V} published the FFD of TRAPPIST-1 using raw data and estimated similar value of the slope but in a slightly different energy range 1.3 $\times$ 10$^{30}$ - 1.2 $\times$ 10$^{33}$ erg. In comparison to other M8 dwarfs in our sample, TRAPPIST-1 has steeper slope than M8.5 dwarf: 2M03264+1919 (-$\alpha$ = -1.5) and shallower slope than another M8 dwarf 2M1221-0843 (-$\alpha$ = -1.8) in slightly different energy ranges. \\ \\
\begin{figure}%
\includegraphics[scale=0.55,angle=0]{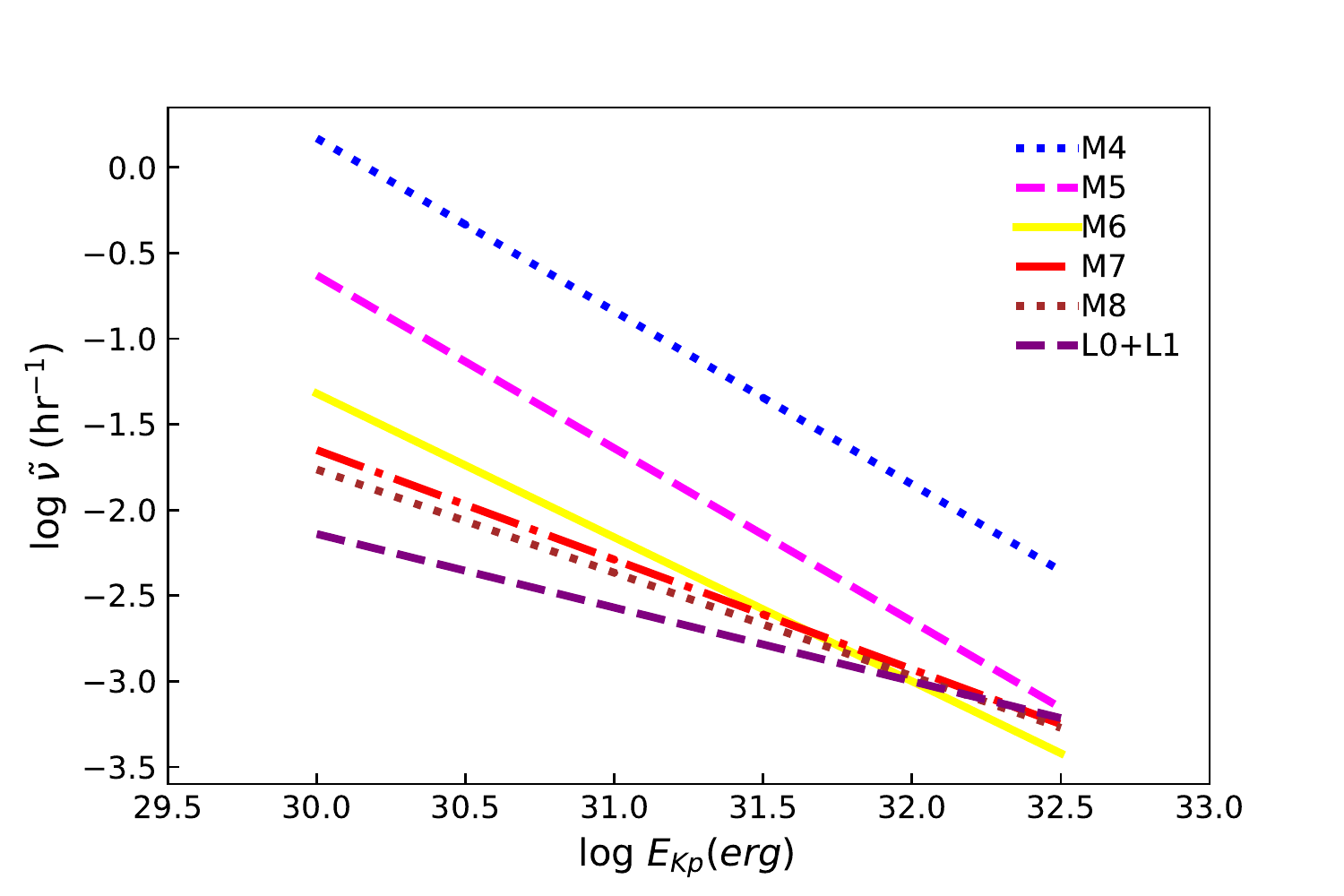} 
\caption{Comparison of average flare rates of each spectral type in our sample with flare rates of GJ1243 (M4) and GJ1245 AB (M5) taken from \cite{2014ApJ...797..121H} and \cite{2015ApJ...800...95L}.}
\end{figure}
\label{fig:compare_hawley}
In Figure \ref{fig:compare_hawley}, we compare the average flare rates of UCDs with a range of spectral types. The average flare rates are computed by mixing both active and less active targets of similar spectral types. For the L0+L1 catagory, we used average flare rate of 2M1232-0951 (L0) and W1906+40 (L1). For comparison with previous works, we compare the flare rates of M4 dwarf GJ 1243 and M5 dwarf GJ 1245 AB taken from \cite{2014ApJ...797..121H} and \cite{2015ApJ...800...95L} who also report the slopes of FFDs using \textit{Kepler} energies of flares. We can see that the flare rates of GJ 1243  are higher than those of late-M dwarfs and early L dwarfs by about two orders of magnitude for flares with energy 10$^{30}$ erg. In addition, Figure \ref{fig:compare_hawley} shows cool stars tend to have shallower slopes. The $\beta$ values of averaged FFDs in Figure \ref{fig:compare_hawley} are 0.84, 0.64, 0.60 and 0.43 for M6, M7, M8 and L0+L1 spectral types respectively.
\begin{figure}
\includegraphics[scale=0.55,angle=0]{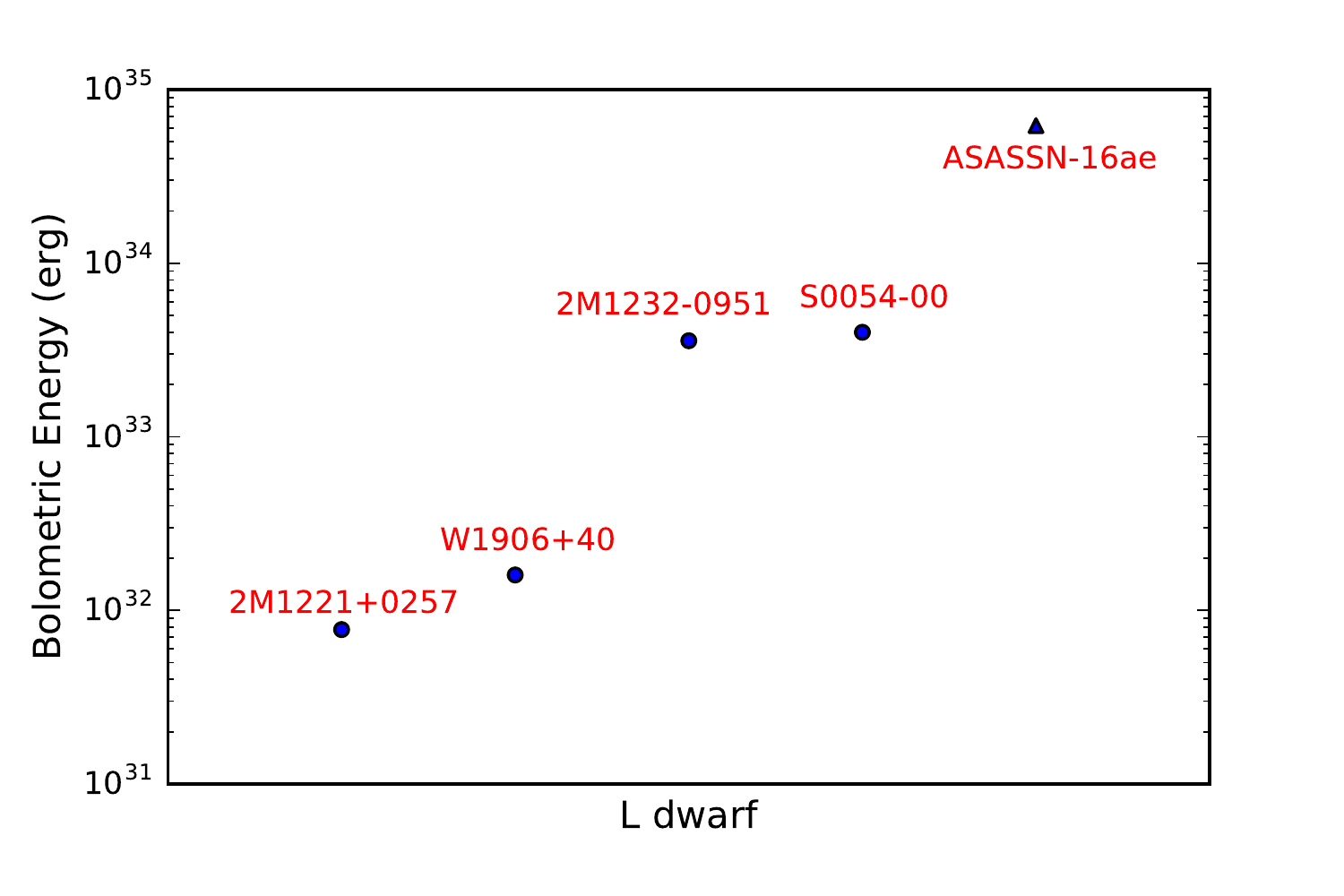} 
\caption{Bolometric energies of biggest flares observed on all five L dwarfs.}
\end{figure}
\label{fig:Ldwarf energies}
\\
\\
The L dwarfs have the lowest flare rates for small energy flares. The FFD of L0 dwarf 2M1232-0951 has a very shallow slope (-$\alpha$ = -1.3) compared to other UCDs in sample. The fitted line covers three orders in magnitude. A shallower slope signifies that occurence rate of bigger flares is higher in this target than other targets in our sample. The FFD of another L dwarf W1906+40 as obtained by \cite{2013ApJ...779..172G} also suggests that L dwarfs have shallower slopes. Figure \ref{fig:compare M7_L0 FFD} compares the FFD of 2M1232-0951 with that of young BD 2M0335+2342. The convergence of two fitted lines at observed high flare energies suggests that both targets have comparable flare rate for such energies. For flare energy log $E_{Kp}$$\sim$29.0 erg, 2M1232-0951 has lower flare rate with the difference being more than one order of magnitude. The convergence of lines is also clearly seen in Figure \ref{fig:compare_hawley} in case of spectral types $\geq$M5, which again indicates comparable flare rates for observed higher flare energies. More fascinating fact about 2M1232-0951 is that it has a flare with the second highest energy among all the targets in our sample. How does it have better efficiency in converting magnetic energy to higher energy flares than those with low energies?  We have to understand the relationship between the magnetic field, volume and rate at which the magnetic field lines are stretched to store energy. The stretching of magnetic field lines may also depend on speed of convective flows.  \\ \\
The steeper slope ($-\alpha$ = -2.0) which we have obtained in case of another L0 dwarf (2M1221+0257) is valid for a very narrow energy range (log \textit{E$_{Kp}$} $\sim$29.7 - 30.4 erg). There is one high energy flare on this target but its energy was not included for fitting purpose. It will be inappropriate to compare the flare rates of two L dwarfs based on the slopes obtained here as the energy ranges considered for obtaining the slopes are very different. As of now, white light flares are observed on five L dwarfs. Two such flaring L dwarfs are discussed in this paper. The remaining three are discussed in \cite{2013ApJ...779..172G}, \cite{2016ApJ...828L..22S} and \cite{2017ApJ...838...22G}. Figure \ref{fig:Ldwarf energies} shows the bolometric (UV/optical/infrared) flare energy distribution of the biggest flares observed on all five L dwarfs. ASASSN-16ae has the largest bolometric flare energy which is $>$ 6.2 $\times$ 10$^{34}$ erg \citep{2016ApJ...828L..22S}. 
\\
\\
The fraction of L0 and L1 dwarfs having chromospheric H$_{\alpha}$ emission is $\sim$90\% and $\sim$67\%, respectively, with a decline in H$_{\alpha}$ activity in comparison to earlier spectral types \citep{2015AJ....149..158S}. This may be due to the lower effective temperatures and hence less ionization reducing the effectiveness of the interaction between the magnetic field and gas \citep{2002ApJ...571..469M}. In addition, the L0-L1 dwarfs do not have clearly developed rotation-activity connections despite being rapid rotators \citep{2008ApJ...684.1390R}. We do not have proper information about the rotation periods, or ages, or activity levels for the two L dwarfs 2M1232-0951 and 2M1221+0257. But what we know is that 2M1232-0951 has a longer time-scale variability \citep{2013MNRAS.428.2824K}, and 2M1221+0257 has a variable H$_{\alpha}$ emission with equivalent width 25.65$\AA$ and log$L_{H\alpha}/L_{bol}$ = -4.18  \citep{2008ApJ...684.1390R}. This limited information is not enough to interpret the observed results of L dwarfs.  One possible physical process which might contribute to shallower FFD slopes in L dwarfs is discussed by \cite{2018ApJ...854...14M}. The model is based on production of flares by instability in coronal magnetic loops in which the footpoints of magnetic flux ropes are subject to random walk due to convective flows. Shallower slopes of the FFD are found to arise in the presence of reduced electrical conductivity in the coolest stars.
\\
\\
We observed superflares on two targets: 2M1232-0951 and 2M0835+1029. Those flares have total bolometric (UV/optical/infrared) energies 3.6 $\times$ 10$^{33}$ erg and 8.9 $\times$ 10$^{33}$ erg. They have very short FWHM time scales of $\sim$2 minutes. In the case of 2M1232-0951, the superflare brightened by a factor of $\sim$144 relative to the quiescent photospheric level. Likewise, the superflare observed on 2M0835+1029 brightened by a factor of $\sim$60 relative to the quiescent photospheric level. Another superflare with bolometric (UV/optical/infrared) energy 2.6 $\times$ 10$^{34}$ was observed on a 130-Myr old brown dwarf CFHT-PL-17 \citep{2017ApJ...845...33G}. We have also detected the most powerful white light flare on another young brown dwarf \citep{2018AAS...23134910P}. The presence of superflares on so many cool targets (the flares discussed here, plus the flares discussed in \cite{2016ApJ...828L..22S} and \cite{2017ApJ...838...22G}) suggests that this is a universal phenomenon which is commonly observed on solar-type stars \citep{2012Natur.485..478M,2013ApJS..209....5S,2013ApJ...771..127N}. It also suggests that very low mass stars and brown dwarfs are magnetically active despite having cool atmospheres. The underlying nature of magnetic dynamo is still unknown in such objects. 
\\
\\
\begin{figure*}
\includegraphics[scale=0.75,angle=0]{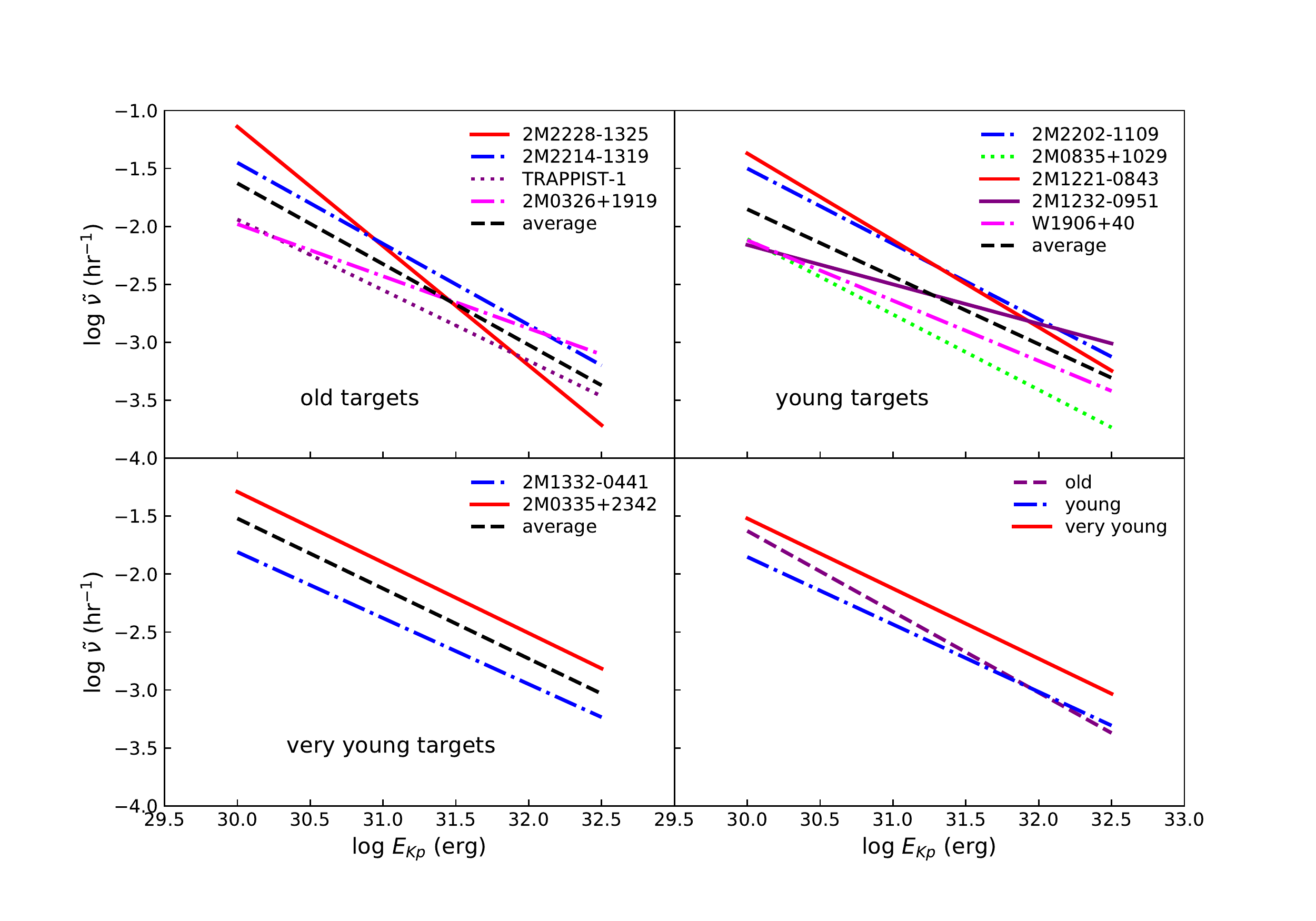} 
\caption{Comparison of flare rates of targets of different kinematic ages. We used our tangential velocity estimates as age indicators. The upper two plots compare flare rates of old and young targets. The lower left plot compares the flare rates of very young targets and the lower right plot compares the average flare rates of old, young and very young targets. The dashed lines in the upper two plots and the lower left plot show the average flare rates of targets included in those plots.}
\end{figure*}
\label{fig:flare rates based on ages}
\begin{table} 
	\centering
	\caption{Average slopes of FFDs of targets
	 classified by kinematic age}
	\begin{tabular}{ccc}
	\hline
	\hline
	Target type & Average V$_{tan}$ & Average $\beta$ \\
	& km s$^{-1}$ &  \\
	\hline
	old & 59 & 0.70 \\
	young & 28 & 0.58  \\
	very young & 10 & 0.61 \\ 
	\hline
	\end{tabular}
\end{table}
\label{table:average flare rates based on ages}
If we make use of the tangential velocity estimates as age indicators, we find that our targets have different ages. TRAPPIST-1 and 2M0326+1919 seem to be the oldest targets and 2M1332-0441 seems to be the youngest one. In support of our claim of a great age for TRAPPIST-1, we may cite \cite{2017ApJ...845..110B} who have reported an age of 7.6$\pm$2.2 Gyr. Despite having the same tangential velocity and spectral type, the FFD of TRAPPIST-1 has a slightly steeper slope than 2M0326+1919: moreover, TRAPPIST-1 has a higher occurence rate of low energy flares than the same M8.5 dwarf. The FFD of 2M2228-1325 has a steeper slope than 2M2202-1109, though 2M2228-1325 has a larger tangential velocity and the same spectral type. Likewise, 2M1332-0441 has a shallower slope than other targets of similar spectral type, despite having the lowest tangential velocity, and is therefore presumably the youngest. In Figure \ref{fig:flare rates based on ages}, we compare the flare rates of targets having similar ages with different spectral types. The targets are categorized as very young, young, or old according to the tangential velocity estimates we have. The average tangential velocities and average slopes of FFDs of the three categories of targets are listed in Table \ref{table:average flare rates based on ages}. 2M1221+0257 is not included for this purpose because its observed flare energies lie in a very narrow range. While there is a slight difference in the slopes of FFDs of targets of various ages in Table \ref{table:average flare rates based on ages}, we cannot conclude anything definitive because of the small sample sizes.\\ 
\\
There is diversity both in age and spectral type of our targets. The diversity in the slopes of FFDs and flare rates observed in our targets suggests that the flare rate of a given target may depend on many factors rather than just age and spectral type (effective temperature). Some of such factors may be rotation rate, magnetic field topology, the number of spots, etc. It will be interesting to compare the flare rates of some targets which will again be observed by \textit{K2} in future campaigns. If we can use new data for the same targets, we may be able to confirm if the flare rates of UCDs remain constant, change slightly or change drastically within short intervals of time. Due to small sample size and smaller number of flares detected on faint targets, we cannot conclude any relation between the spectral type or age and the slope -$\alpha$ of FFD. We need to study the FFDs for a larger sample of UCDs over a wide range of energies to see if any such relations exist. The results of \cite{2016ApJ...831..131L} suggests that the occurence rate (\textit{R}) of big solar flares and front-side halo coronal mass ejections are higher during the descending phase of solar cycle than other phases. They report a strong anti-correlation between \textit{R} and annual average latitude of sunspot groups. Is this just a coincidence or due to some underlying physical phenomenon? We may also think of same scenario in case of our targets. \cite{2016ApJ...830L..27R} has reported some evidence of magnetic cycles in UCDs. The flare rate on UCDs might depend on the phase of their magnetic cycle. More information about rotational velocities of UCDs will be helpful to understand such correlations (if they exist). \\
 \\
The FFDs of two of our targets seem to show deviations from a single power law dependence. Using the likelihood ratio test, we found that the FFD of one target (2M2228-1325) seems to follow a broken power law distribution while the FFD of another target (2M0326+1919) seems to follow a power law distribution with exponential cutoff. Unfortunately, since the number of flares observed on both targets is small, we cannot conclude if such deviations of FFDs from regular power laws are due to instrumental sensitivity or due to saturation at large energies. \cite{2005stam.book.....G}  (p. 227) mentions that the curvatures seen in FFD of some targets were absent when they were observed again, and the number of observed flares was increased. Curved FFDs can be seen in case of EQ Peg, UV Cet and AD Leo in Figure 38 (p. 224) of \cite{2005stam.book.....G}. Other examples of departures of FFD from single power-laws are provided by GJ 1243 and GJ 1245 AB \citep{2014ApJ...797..121H}, and by KIC 11551430 \citep{2016ApJ...829...23D}.\\
\\
The flare rates of UCDs reported in this paper should be helpful in predicting the number of flares on targets of similar types, which could be observed by future photometric surveys. They are also very important for gyrochronology and studying the planets in habitable zone of stars like TRAPPIST-1. The biggest flares might be capable of damaging atmospheric chemistry and other habitable conditions of the planets. However, detailed discussion about the possible impact of flares of TRAPPIST-1 on its planetary atmospheres is beyond the scope of this paper. One can find the necessary information in recent papers like \cite{2017ApJ...851...77R,2017arXiv171108484T}, etc.\\ 

\acknowledgements{Acknowledgements}\\ 
We thank Rachel Osten, Eric D. Feigelson and James R. A. Davenport for their valuable suggestions on some methods and statistical tools used for flare analysis and FFD fitting. The material is based upon work supported by NASA under award Nos. NNX15AV64G, NNX16AE55G and NNX16AJ22G. P.K.G.W. and E.B. acknowledge support for this work from the National Science Foundation through Grant AST-1614770. Some/all of the data presented in this paper were obtained from the Mikulski Archive for Space Telescopes (MAST). STScI is operated by the Association of Universities for Research in Astronomy, Inc., under NASA contract NAS5-26555. Support for MAST for non-HST data is provided by the NASA Office of Space Science via grant NNX09AF08G and by other grants and contracts. This paper includes data collected by the Kepler mission. Funding for the Kepler mission is provided by the NASA Science Mission directorate. The Pan-STARRS1 Surveys (PS1) and the PS1 public science archive have been made possible through contributions by the Institute for Astronomy, the University of Hawaii, the Pan-STARRS Project Office, the Max-Planck Society and its participating institutes, the Max Planck Institute for Astronomy, Heidelberg and the Max Planck Institute for Extraterrestrial Physics, Garching, The Johns Hopkins University, Durham University, the University of Edinburgh, the Queen's University Belfast, the Harvard-Smithsonian Center for Astrophysics, the Las Cumbres Observatory Global Telescope Network Incorporated, the National Central University of Taiwan, the Space Telescope Science Institute, the National Aeronautics and Space Administration under Grant No. NNX08AR22G issued through the Planetary Science Division of the NASA Science Mission Directorate, the National Science Foundation Grant No. AST-1238877, the University of Maryland, Eotvos Lorand University (ELTE), the Los Alamos National Laboratory, and the Gordon and Betty Moore Foundation.

\vspace{5mm}
\facilities{\textit{Kepler}}

\software{astropy \citep{2013A&A...558A..33A}, emcee \citep{2013PASP..125..306F}}

\bibliographystyle{aasjournal}
\bibliography{/Users/rishipaudel/GoogleDrive/Research/astrobib}

\end{document}